\documentclass[onecolumn,twoside]{IEEEtran}
\usepackage{amsthm}
\usepackage{times,amssymb,amsmath,amsfonts,float,nicefrac,color,bbm,mathrsfs,caption,float}
\usepackage{algorithm,enumerate,multirow,caption,tikz,graphicx}
\usepackage[mathscr]{eucal}
\usepackage{sidecap}
\usepackage{algpseudocode}
\usepackage{verbatim}
\usepackage{epstopdf}
\usepackage{textcomp}
\usepackage{bm}
\usepackage{cases}
\usepackage{multirow,makecell}

\usepackage{stmaryrd}
\usepackage{mathabx}
\usepackage[noadjust]{cite}
\usepackage{booktabs}
\usepackage[normalem]{ulem}
\usepackage[top=1in,bottom=1in,left=1in,right=1in]{geometry}
\allowdisplaybreaks

\newcommand{\RN}[1]{%
  \textup{\expandafter{\romannumeral#1}}%
}

\newcommand\remove[1]{}

\allowdisplaybreaks

\newtheorem{theorem}{Theorem}

\newtheorem{lemma}{Lemma}
\newtheorem{corollary}{Corollary}

\newtheorem{example}{Example}
\newtheorem{remark}{Remark}

\newcommand{\Fq}{\mathbb{F}_{q}}
\newcommand{\Fqm}{\mathbb{F}_{q^m}}

\newcommand{\ff}{{\mathbb F}}

\newcommand{\bC}{\mathbb{C}}

\newcommand{\bF}{\mathbb{F}}

\newcommand{\cC}{\mathcal{C}}

\DeclareMathOperator{\rank}{rank}

\DeclareMathOperator{\lcm}{lcm}
\DeclareMathOperator{\ord}{ord}

\begin{document}
\title{A Class of Narrow-Sense BCH Codes }

\author{\IEEEauthorblockN{Shixin Zhu,}  \and \IEEEauthorblockN{Zhonghua Sun,} \and \IEEEauthorblockN{and Xiaoshan Kai} }

\maketitle
{\renewcommand{\thefootnote}{}\footnotetext{

\vspace{-.2in}

\noindent\rule{1.5in}{.4pt}

{The authors are with the School of Mathematics, Hefei University of Technology, Hefei 230009, China. Their research is supported by the National Natural Science Foundation of China under Grants 61772168 and 61572168. Emails: zhushixin@hfut.edu.cn; sunzhonghuas@163.com; kxs6@sina.com. }}
}
\renewcommand{\thefootnote}{\arabic{footnote}}
\setcounter{footnote}{0}

\begin{abstract}
    BCH codes are an important class of cyclic codes which have applications in satellite communications, DVDs, disk drives, and two-dimensional bar codes. Although BCH codes have been widely studied, their parameters are known for only a few special classes. Recently, Ding et al. made some new progress in BCH codes. However, we still have very limited knowledge on the dimension of BCH codes, not to mention the weight distribution of BCH codes. In this paper, we generalize the results on BCH codes from several previous papers.
\begin{itemize}
    \item [(\romannumeral1)]The dimension of narrow-sense BCH codes of length $\frac{q^m-1}{\lambda}$ with designed distance $2\leq \delta \leq \frac{q^{\lceil(m+1)/2 \rceil}-1}\lambda+1$ is settled, where $\lambda$ is any factor of $q-1$.
    \item [(\romannumeral2)] The weight distributions of two classes of narrow-sense BCH codes of length $\frac{q^m-1}2$ with designed distance $\delta=\frac{(q-1)q^{m-1}-q^{\lfloor(m-1)/2\rfloor}-1}2$ and $\delta=\frac{(q-1)q^{m-1}-q^{\lfloor(m+1)/2\rfloor}-1}2$ are determined.
    \item[(\romannumeral3)] The weight distribution of a class of BCH codes of length $\frac{q^m-1}{q-1}$ is determined. 
\end{itemize}
In particular, a subclass of this class of BCH codes is optimal with respect to the Griesmer bound. Some optimal linear codes obtained from this class of BCH codes are characterized.
\end{abstract}

\noindent{\it Keywords: Cyclic codes, BCH codes, Weight distribution}

\section{Introduction}\label{section:parameters bch}

\subsection{Backgrounds}

  Let $q$ be a prime power and $\ff_q$ be the finite field with $q$ elements. Let $n,k$ be positive integers with $1\leq k\leq n$. An $[n,k]$ linear code $\mathcal{C}$ is a subspace of the vector space $\Fq^n$ with dimension $k$. If this linear code $\cC$ is, in addition, closed under the cyclic shift, i.e., $(c_{n-1},c_0,c_1,\ldots,c_{n-2})\in \cC$ for any $(c_0,c_1,\ldots,c_{n-1})\in \cC$, then $\cC$ is called a cyclic code. Each vector $\bm{c}=(c_0,c_1,\ldots,c_{n-1})$ is customarily identified with its polynomial representation $c(x)=c_0+c_1x+\dots+c_{n-1}x^{n-1}$, and a code is identified with the set of polynomial representations of its codewords. A linear code $\cC$ of length $n$ over $\bF_q$ is cyclic if and only if $\cC$ is an ideal of $\Fq[x]/\langle x^n-1\rangle$. It is well known that every ideal of $\Fq [x]/\langle x^n-1\rangle$ is principal. Hence, there is a monic divisor $g(x)$ of $x^n-1$ such that $\cC=\langle g(x)\rangle$. The polynomial $g(x)$ is called the generator polynomial of $\cC$, and $h(x)=\frac{x^n-1}{g(x)}$ is called the parity-check polynomial of $\cC$. If $h(x)$ has $t$ irreducible factors over $\ff_q$, we say such a cyclic code $\cC$ has $t$ nonzeros.

  Suppose $n$ is a positive integer with $\gcd(n,q)=1$. Let $m=\ord_n(q)$, i.e., the multiplicative order of $q$ modulo $n$ is $m$, and $\alpha$ be a primitive element in $\ff_{q^m}$. Assume that $q^m-1=n\lambda$ and $\theta=\alpha^\lambda$, then $\theta$ is a primitive $n$-th root of unity. For each $0\leq i\leq n-1$, let $m_i(x)$ be the minimum polynomial of $\theta^i$ over $\ff_q$. A cyclic code of length $n$ over $\ff_q$ is called a BCH code with designed distance $\delta$ if its generator polynomial is of the form
  \[\begin{split}
  \lcm (m_{b}(x),m_{b+1}(x),\ldots, m_{b+\delta-2}(x)),
  \end{split}\]
   where lcm denotes the least common multiple of the polynomials, $2\leq \delta\leq n$ and $b\geq 0$. Denote such a BCH code with designed distance $\delta$ by $\cC_{(q,m,\lambda,\delta,b)}$. If $b=1$ it is called a narrow-sense BCH code and we denote it by $\cC_{(q,m,\lambda,\delta)}$. Clearly, $\cC_{(q,m,\lambda,\delta+1,0)}\subseteq \cC_{(q,m,\lambda,\delta)}$. We denote $\cC_{(q,m,\lambda+1,0)}$ by $\widehat{\cC}_{(q,m,\lambda,\delta)}$.

  BCH codes were invented by Hocquenghem \cite{hocquenghem1959codes}, and independently by Bose and Ray-Chaudhuri \cite{bose1960class}. One of the key features of BCH codes is a precise control over the number of symbol errors correctable by the code. Another advantage of BCH codes is that they have efficient encoding and decoding algorithms. Due to BCH codes have such good properties, they are widely used in DVDs, solid-state drives, compact disc players, disk drives, two-dimensional bar codes and satellite communications.

\subsection{Known Results}

BCH codes have been extensively studied in the literature (\cite{DY1996codes,Desaki1997weight,augot1992locator,augot1994idempotents,berlekamp1967,bose1960class,charpin1990,charpin1998open,charpin2006,ding2015bose,ding2015parameters,ding2017dimension,helgert1973,hocquenghem1959codes,kasami1967minimum,kasami1969,kasami1969remarks,kasami1972minimum,li2016narrow,li2017lcd,li2017two,macwilliams1977theory,yan2017,yue1992structure,liu2017dimensions,mann1962,fujiwara1986,geer1994,krasikov2001,keren1999,kasami1985,li2017,li2017minimum,lin1967,mandelbaum1980,wolf1969,yue2000}). Nonetheless, their parameters are known for only a few special classes. As pointed out by Charpin
\cite{charpin1998open}, the dimension and minimum distance of BCH codes are difficult to determine in general. The dimensions of the BCH codes $\cC_{(q,m,\lambda,\delta)}$ were investigated in a lot of papers. We roughly list them in the Table \ref{table:13}. Besides the results in Table \ref{table:13}, for $q^{\lfloor m/2 \rfloor}<n\leq q^m-1$ and $2\leq\delta \leq \frac{nq^{\lceil m/2\rceil}}{q^m-1}$, the dimension of $\cC_{(q,m,\lambda,\delta)}$ was settled by Aly et al. \cite{aly2007quantum}. Recently, the dimensions of some BCH codes $\cC_{(q,m,\lambda,\delta,b)}$ with $b\neq 0,1$ were settled in \cite{li2017lcd,liu2017dimensions,li2017two}. 

\begin{table}
	\renewcommand\arraystretch{1.4}
	\caption{KNOWN RESULTS ON DIMENSION OF $\cC_{(q,m,\lambda,\delta)}$}
	\begin{center}
		\begin{tabular}{|c|c|c|}
			\hline
			$\lambda$ & $\delta$& Reference\\
			\hline
          \multirowcell{5}{$\lambda=1$}&$\delta=q^t$&\cite{mann1962}\\\cline{2-3}
            &$\delta=q^{m-2}+1$&\cite{ding2015bose}\\\cline{2-3}
            &$2\leq \delta \leq q^{\lceil m/2 \rceil}+1$;&\multirowcell{2}{\cite{DY1996codes}}\\
            &$q^{m/2}+2\leq \delta \leq 2q^{m/2}+1$, $m$ even&\\\cline{2-3}
            &$2\leq \delta \leq q^{\lceil m/2\rceil+1}$&\cite{liu2017dimensions}\\
			\hline
			\multirowcell{2}{$\lambda=q-1$}&$2\leq \delta \leq q^{m/2}$, $m$ even &\cite{li2017lcd}\\\cline{2-3}
			&$2\leq \delta \leq q^{(m+1)/2}$, $m$ odd & \cite{liu2017dimensions}\\
			\hline
			\multirowcell{4}{$\lambda=q^{\ell}-1$,\\
			$m=2\ell$}&$3\leq \delta\leq q^{\lfloor(\ell-1)/2\rfloor}+2$& \cite{li2017lcd}\\\cline{2-3}
			&$2\leq \delta\leq q^{\lfloor(\ell+1)/2\rfloor}+1$& \cite{liu2017dimensions}\\\cline{2-3}
			&$2\leq \delta\leq 2q^{\ell/2}+3$, $\ell$ even;& \multirowcell{2}{\cite{li2017}}\\
			&$2\leq \delta\leq 2q^{(\ell+1)/2}+2q$, $\ell$ odd.& \\
			\hline
		\end{tabular}
	\end{center}
	\label{table:13}
\end{table}

The exact minimum distance of BCH codes  has been studied in many literatures ( \cite{charpin1990,ding2015bose,ding2015parameters,ding2017dimension,kasami1969,li2016narrow,Desaki1997weight,geer1994,li2017minimum}). The reader is referred to \cite{ding2015bose} for a recent summary of various results on minimum distance of BCH codes. In general, the problem of determining the weight distribution of BCH codes is very difficult, and it is known for only a few special classes. Not much work has been done on determining the weight distribution of BCH codes. We list them in the following two cases.

\begin{itemize}
    \item [(\romannumeral1)] Case 1: $\lambda=1$. For $\delta=(q-1)q^{m-1}-q^{\lfloor(m-1)/2\rfloor}-1$
    and $\delta=(q-1)q^{m-1}-q^{\lfloor(m+1)/2 \rfloor}-1$, when $q=2$, the weight distribution of $\cC_{(q,m,\lambda,\delta)}$ was
    settled by Kasami \cite{kasami1969}; when $q$ is a prime, the weight distribution of $\cC_{(q,m,\lambda,\delta)}$ and $\widehat{\cC}_{(q,m,\lambda,\delta)}$ was settled by Ding et al. \cite{ding2017dimension}. For $\delta=q^3-q^2-q-2$ and $m=3$, the weight distribution of $\cC_{(q,m,\lambda,\delta)}$ was determined by Yan \cite{yan2017}. Recently, For $\delta=q^m-q^{m-1}-q^i-1$, where $\frac{m-2}2\leq i\leq m-\lfloor \frac{m}3\rfloor -1$, the weight distribution of $\widehat{\cC}_{(q,m,\lambda,\delta_i)}$ was determined by Li \cite{li2017minimum}.
    \item[(\romannumeral2)] Case 2: $\lambda=2$ and $q=3$. For $\delta_i=3^{m-1}-1-\frac{3^{\lfloor(m+2i-3)/2\rfloor}-1}2$,
    where $1\leq i\leq 2$, the weight distribution of  $\cC_{(q,m,\lambda,\delta_i)}$ and $\widehat{\cC}_{(q,m,\lambda,\delta_i)}$ was settled by Li et al. \cite{li2016narrow}.
\end{itemize}

\subsection{The contribution of the present paper}

  The objective of this paper is to study narrow-sense BCH codes over $\mathbb{F}_{q}$ of length $\frac{q^m-1}\lambda$, where $\lambda$ is a positive factor of $q-1$. The main contributions are the following:
  \begin{itemize}
  	\item[(\romannumeral1)] For $2\leq \delta \leq\frac{q^{\lceil(m+1)/2 \rceil}-1}\lambda+1$, the dimension of the BCH code $\cC_{(q,m,\lambda,\delta)}$ is completely determined. These results generalize those from \cite{li2017lcd,liu2017dimensions}.
  	\item[(\romannumeral2)] For $\lambda=2$ and $\delta_i=\frac{(q-1)q^{m-1}-q^{\lfloor(m+2i-3)/2\rfloor}-1}{2}$ with $i=1,2$, we give a trace representation for the codewords in $\mathcal{C}_{(q,m,2,\delta_i)}$ and $\widehat{\mathcal{C}}_{(q,m,2,\delta_i)}$. By using exponential sums, the weight distribution of the BCH code $\cC_{(q,m,\lambda,\delta_i)}$ and $\widehat{\cC}_{(q,m,\lambda,\delta_i)}$ is settled. These results generalize those from \cite{li2016narrow}.
  	\item[(\romannumeral3)] For $m=a(q-1)+1$ or $a(q-1)+2$ for some integer $a\geq 1$, the first largest $q$-cyclotomic coset leader modulo $\frac{q^m-1}{q-1}$ is determined, and then the weight distribution of a class of BCH codes of length $\frac{q^m-1}{q-1}$ is determined.
  \end{itemize}

 The paper is organized as follows. In Section \ref{sec2}, we give some background and recall some basic results on character sums. By using cyclotomic cosets, the dimension of this class of narrow-sense BCH codes is determined in Section \ref{sec3}. In Section \ref{sec4}, we find a trace representation for the codewords in $\mathcal{C}_{(q,m,2,\delta_i)}$ and $\widehat{\mathcal{C}}_{(q,m,2,\delta_i)}$, where
$\delta_i=\frac{(q-1)q^{m-1}-q^{\lfloor(m+2i-3)/2\rfloor}-1}{2}$ with $i=1,2$. In addition, by using exponential sums and the theory of quadratic forms over finite fields, the weight distributions of $\mathcal{C}_{(q,m,2,\delta_i)}$ and $\widehat{\mathcal{C}}_{(q,m,2,\delta_i)}$ are determined. Moreover, the weight distribution of a class of BCH codes of length $\frac{q^m-1}{q-1}$ is also determined. Furthermore, a subclass of such BCH codes meeting the Griesmer bound is presented. Compared with the table of the best known linear codes maintained by Markus Grassl at http://www.codetables.de/, which is called the Database later in this paper, these two classes of BCH codes are sometimes among the best liner codes known. Finally, the conclusion of the paper is given in Section \ref{sec5}.

\section{preliminaries}\label{sec2}

Throughout this paper, let $\lambda$ be a positive divisor of $q-1$ and $n=\frac{q^{m}-1}\lambda$, where $m\geq 2$ is a positive integer. Clearly, $\gcd(n,q)=1$ and $\ord_{n}(q)=m$.

 
 Let $\alpha$ be a primitive element of $\Fqm$ and put $\theta=\alpha^\lambda$, then $\theta$ is a primitive $n$-th root of unity. For any $0\leq i\leq n-1$, the $q$-cyclotomic coset of $i$ modulo $n$ is defined as $\bC_{i}=\left\lbrace iq^j~(\text{mod}~n):0\leq j\leq l_i-1\right\rbrace $, where $l_i$ is the least positive integer such that $iq^{l_i}\equiv i~(\text{mod}~n)$ and is the size of $\bC_i$. Obviously, $l_i\mid m$. The smallest element in $\bC_i$ is called the coset leader of $\bC_i$. For every $2\leq \delta\leq n$ and $b\geq0$, we define
\[\begin{split}
g_{(q,m,\lambda,\delta,b)}(x)=\prod_{z\in \mathcal{D}}(x-\theta^z),~\text{where}~\mathcal{D}=\bigcup_{j=0}^{\delta-2} \bC_{b+j}.
\end{split}\]
Obviously, $g_{(q,m,\lambda,\delta,b)}$ is the generator polynomial of $\cC_{(q,m,\lambda,\delta,b)}$. If $b=1$, the dimension of $\cC_{(q,m,\lambda,\delta)}$ is
\[\begin{split}
\dim(\cC_{(q,m,\lambda,\delta)})=n-|\bigcup_{i=1}^{\delta-1}\mathbb{C}_i|.
\end{split}\]
Moreover, $\dim(\widehat{\cC}_{(q,m,\lambda,\delta)})=\dim(\cC_{(q,m,\lambda,\delta)})-1$. The following is the well known BCH bound.

\begin{lemma}\cite[Ch. 7, Th. 8]{macwilliams1977theory} \label{l1}
   The minimum distance of $\cC_{(q,m,\lambda,\delta,b)}$ is at least $\delta$.
\end{lemma}

   Let $p$ be the characteristic of $\Fq$, then $q$ is a power of $p$. Let ${\rm Tr}_q^{q^m}$ be the trace mapping from $\Fqm$ to $\Fq$ and $\zeta_p=e^{\frac{2\pi i}p}$, where $m$ is a positive integer. For any given $a\in \Fq$, the function $\chi_a(x)=\zeta_p^{{\rm Tr}_p^q(ax)}$ is an additive character of $\Fq$. The character $\chi_1$ is called the canonical character of $\Fq$. Let $\beta$ be a fixed primitive element of $\Fq$. For each $0\leq j\leq q-2$, the function $\psi_j$ with $\psi_j(\beta^k)=\zeta_{q-1}^{jk}$ for $0\leq k\leq q-2$ defines a multiplicative character of $\Fq$, and every multiplicative character of $\Fq$ can be defined in this way. The character $\psi_0$ is called the trivial multiplicative character of $\Fq$. When $q$ is odd, the character $\psi_{\frac{q-1}2}$ is called the quadratic character of $\Fq$, and is usually denoted by $\eta$. Let $\psi$ be a multiplicative character and $\chi$ an additive character of $\Fq$. Then the Gaussian sum $G(\psi,\chi)$ is defined by $G(\psi,\chi)=\sum_{x\in\Fq^*}\psi(x)\chi(x)$. From now on we shall denote the Gaussian sum $G(\eta,\chi_1)$ over $\ff_q$ by $G_q$. The explicit value of $G_q$ is known.

\begin{lemma}\cite[Theorems 5.15, 5.33]{lidl1977fields}\label{l2}
	Let $q=p^s$, where $p$ is an odd prime and $s$ is a positive integer. Then
	\[\begin{split}
	G_q=\begin{cases}(-1)^{s-1}\sqrt{q} ~&{\rm if}~p\equiv 1~({\rm mod}~4),\\
	(-1)^{s-1}(\sqrt{-1})^s\sqrt{q} ~&{\rm if}~p\equiv 3~({\rm mod}~4),\\
	\end{cases}
	\end{split}\]
    and for each $a\in \Fq^*$,
    \[\begin{split}
    \sum_{x\in \Fq^*}\zeta_p^{{\rm Tr}_p^q(ax^2)}=\eta(a)G_q-1,
    \end{split}\] 
    where $\eta$ is the quadratic character of $\Fq$.
\end{lemma}

    We recall the following trace representation of cyclic codes, which is a direct consequence of Delsarte's Theorem \cite{Delsarte1975On}.

\begin{lemma}\cite[Proposition 18]{li2016narrow}\label{l3}
    Let $q$ be a prime power and $m=\ord_n(q)$. Let $\theta$ be a primitive $n$-th root of unity in $\Fqm$ and $\cC$ be a cyclic code of length $n$ over $\Fq$. Suppose $\cC$ has $t$ nonzeros and let $\theta^{i_1},\theta^{i_2},\ldots,\theta^{i_t}$ be the $t$ roots of its parity-check polynomial which are not conjugate with each other. Denote the size of the $q$-cyclotomic coset $\bC_{i_j}$ to be $m_j$, $1\leq j\leq t$. Then $\cC$ has the following trace representation
    \[\begin{split}
    \cC=\left\lbrace c(a_1,a_2,\ldots,a_t): a_j\in \ff_{q^{m_j}},~1\leq j\leq t\right\rbrace,
    \end{split}\]
    where $c(a_1,a_2,\ldots, a_t)=\left(\sum_{j=1}^t{\rm Tr}^{q^{m_j}}_q(a_j\theta^{-\ell i_j}) \right)_{\ell=0}^{n-1}$.
\end{lemma}

We give a brief introduction to the theory of quadratic forms over
finite fields, which is used to calculate the weight distribution of
BCH codes. Quadratic forms have been well studied (\cite{feng2008weight,luo2008weight,luo2009exponential,wan2002geometry,zhou2014class}).
The form is called a quadratic form over $\Fq$ if is a homogeneous
polynomial of degree two in the form
\[\begin{split}
Q(x_1,x_2,\ldots,x_m)=\sum_{1\leq i\leq j\leq m}a_{ij} x_ix_j,~a_{ij}\in \Fq.
\end{split}\]
If $q$ is odd, for a quadratic form $Q(x_1,x_2,\ldots,x_m)$ in $m$
variables over $\Fq$, there exists a symmetric matrix $A$ of order
$m$ over $\Fq$ such that $Q(x)=\boldsymbol{x}A\boldsymbol{x}'$, where
$\boldsymbol{x}=(x_0,x_1,\ldots,x_{m-1})\in \Fq^m$ and $\boldsymbol{x}'$ denotes the
transpose of $\boldsymbol{x}$. Let $r=\rank A$, then there exists $M\in{\rm GL}_m(\Fq) $ such that $B=MAM'$ is a diagonal matrix and
$B={\rm diag}(a_1,a_2,\ldots,a_r,0,\ldots,0)$, where $a_i\in
\Fq^*$. Let $\triangledown=a_1a_2\cdots a_r$ and assume that
$\triangledown=1$ when $r=0$. Let $\eta$ be the quadratic character
of $\Fq$, then $\eta(\triangledown)$ is an invariant of $M$ under
the conjugate action of $M\in {\rm GL}_m(\Fq) $. We identify $\Fqm$
with the $m$-dimensional $\Fq$-vector space. The following results
are useful in the sequel.

\begin{lemma}\cite[Lemma 1]{luo2008weight} \label{l4}
	Let $q$ be an odd prime power and $Q(x)$ be a quadratic form in $m$ variables of rank $r$ over $\Fq$. Then
    \[\begin{split}
     \sum_{x\in \Fqm}\zeta_p^{{\rm Tr}_p^q(Q(x))}=\begin{cases}
    \pm q^{m-\frac{r}2}&{\rm if}~ q\equiv 1({\rm mod}~4),\\
    \pm (\sqrt{-1} )^r q^{m-\frac{r}2}&{\rm if}~q\equiv 3({\rm mod}~4).
    \end{cases}
    \end{split}\]
    The following identity holds (see \cite[Lemma 9]{li2016narrow}): 
    \[\begin{split}
    \sum_{x\in \Fqm}\zeta_p^{{\rm Tr}_p^q(yQ(x))}=\eta(y^r)\sum_{x\in \Fqm}\zeta_p^{{\rm Tr}_p^q(Q(x))},~\forall y \in \bF_q^*,
    \end{split}\]
    where $\eta$ is the quadratic character of $\Fq$.
\end{lemma}

\section{The dimension of BCH code of length $n=\frac{q^m-1}\lambda$} \label{sec3}

In this section, we will determine the dimension of the BCH codes $\cC_{(q,m,\lambda,\delta)}$ for $\delta-1\leq
\frac{q^{\left\lceil\frac{m+1}2 \right\rceil}-1}\lambda$. Recall
\[\begin{split}
\dim(\cC_{(q,m,\lambda,\delta)})=n-|\bigcup_{i=1}^{\delta-1}\bC_i|.
\end{split}\] 
Let $\Gamma_1=\{i: 1\leq i\leq \delta-1 \text{ and }i\not\equiv 0~(\text{mod}~q)\}$. Then $\dim(\cC_{(q,m,\lambda,\delta)})=n-|\bigcup_{i\in \Gamma_1}\bC_i|$, since if $i\equiv 0~(\text{mod}~q)$ there exists an integer $j$ with $1\leq j<i$ such that $\bC_j=\bC_i$. Let $\Gamma_2$ denote the set of coset leaders in $\Gamma_1$ and $\Gamma_3$ the set of non coset leader in $\Gamma_1$. Then $\Gamma_1=\Gamma_2\bigcup \Gamma_3$. Note that if $i\in \Gamma_3$, there is an integer $1\leq j<i$ such that $j\in \bC_i$ and $j$ is a coset leader of $\bC_i$. That is, for every $i\in \Gamma_3$, there exists an integer $j\in \Gamma_2$ such that $\bC_i=\bC_j$. It follows that
\[\begin{split}
\dim(\cC_{(q,m,\lambda,\delta)})&=n-\sum_{i\in \Gamma_2}|\bC_i|\\
&=n-\sum_{i\in \Gamma_1}|\bC_i|+\sum_{i\in \Gamma_3}|\bC_i|.
\end{split}\]
Hence, to determine the dimension of the code $\cC_{(q,m,\lambda,\delta)}$, we need to find out the coset leader of $\bC_i$ and its cardinality for each $i\in \Gamma_1$. 

The following result given in \cite{aly2007quantum} will be useful for determining coset leaders when $\delta$ is small.

\begin{lemma}\cite[Lemmas 8, 9]{aly2007quantum}\label{l5}
    Let $n$ be an integer with $q^{\lfloor \frac{m}2\rfloor}<n\leq q^m-1$, where $m=\ord_n(q)$. Then the $q$-cyclotomic coset $\bC_i$ has cardinality $m$ for all $i$ in the range $1\leq i \leq \frac{nq^{\lceil \frac{m}2 \rceil}}{q^m-1}$. Moreover, the following assert holds: every $i$ with $i\not\equiv 0~({\rm mod}~q)$ in this range is a $q$-cyclotomic coset leader modulo $n$.
\end{lemma}

When $m$ is odd, by Lemma \ref{l5}, we have the following conclusion.

\begin{theorem}\label{th1}
	Let $m\geq 3$ be odd. For every integer $\delta$ with $1\leq \delta-1\leq \frac{q^{\frac{m+1}2}-1}\lambda$, $\mathcal{C}_{(q,m,\lambda,\delta)}$ has length $n=\frac{q^m-1}\lambda$, minimum distance $d\geq \delta$ and dimension $n-m\left\lceil \frac{(\delta-1)(q-1)}{q} \right\rceil $.
\end{theorem}

Now we consider the dimension of $\cC_{(q,m,\lambda,\delta)}$ when
$m=2h\geq 4$ and $\delta-1\leq \frac{q^{h+1}-1}\lambda$. When $\lambda=1$, the following result was prove in \cite{DY1996codes,yue1992structure,liu2017dimensions,aly2007quantum}.

\begin{lemma}\label{l6}
	 Let $m=2h\geq 4$. Let $i$ be an integer with $1\leq i\leq \frac{q^{h+1}-1}{\lambda}$ and $i\not\equiv 0 ~({\rm mod}~q)$. Let $f(a,b,c)=aq^h+\frac{b(q^h-1)}\lambda+c$, where $a,b,c$ are integers. Then $i$ is not a $q$-cyclotomic coset leader modulo $n$ if and only if $i\in\Delta_0\bigcup \Delta_1\bigcup \Delta_2$, where 
	 \[\begin{split}
	 &\Delta_0=\left\lbrace f(a,0,c) : 1\leq c<a\leq \frac{q-1}\lambda\right\rbrace,\\
	 \Delta_1=&\left\lbrace f(a,b,c): 1\leq c\leq a<\frac{q-1}\lambda,1\leq b<
	 \lambda\right\rbrace,\\
     \Delta_2=&\left\lbrace f(a,b,a+1): 0\leq a< \frac{q-1}\lambda,\frac{\lambda}2<b<\lambda\right\rbrace.
	 \end{split}\]
\end{lemma}

\begin{IEEEproof}
      We claim that an integer $1\leq i<n$ is not the coset leader in the $q$-cyclotomic coset of $i$ modulo $n$ if and only if $\lambda i$ is not the coset leader in the $q$-cyclotomic coset of $\lambda i$ modulo $\lambda n$. In fact, $i$ is not a coset leader if and only if there exists an integer $j$ with $1\leq j<i$ such that $i\equiv j q^s~(\text{mod}~n)$, for some integer $s$. Note that $i\equiv j q^s ~(\text{mod}~n)$ is equivalent to $\lambda i\equiv \lambda j q^s ~(\text{mod}~\lambda n)$. Hence, the above assert holds.
      
      We divide $\lambda$ into two cases to prove our result.
      \begin{itemize}
      	\item [(\romannumeral1)] If $\lambda=1$, an integer $i$ with $1\leq i\leq q^{h+1}-1$ and $i\not\equiv 0 ~(\text{mod}~q)$ is not a coset leader if and only if $i=aq^h+c$, where $1\leq c<a\leq q-1$, which has been proven in \cite{liu2017dimensions}.
      	\item [(\romannumeral2)] If $\lambda\geq 2$, an integer $i$ with $1\leq i\leq \frac{q^{h+1}-1}\lambda$ and $i\not\equiv 0 ~(\text{mod}~q)$ is not the coset leader in the $q$-cyclotomic coset of $i$ modulo $n$ if and only if $\lambda i$ is not the coset leader in the $q$-cyclotomic coset of $\lambda i$ modulo $\lambda n$. From Case (\romannumeral1), $\lambda i=i_hq^h+i_0$ for integers $1\leq i_0<i_h\leq q-1$.
      	\item [(a)] If $\lambda\mid i_h$, then $\lambda\mid i_0$. Suppose $i_h=\lambda a$ and $i_0=\lambda c$. Then, $i=aq^h+c$, where $1\leq c<a\leq \frac{q-1}\lambda$. That is, $i\in \Delta_0$.
      	\item [(b)] If $\lambda\nmid i_h$, there exist integers $a,b$ such that $i_h=\lambda a+b$, where $1\leq b<\lambda$ and $0\leq a<\frac{q-1}\lambda$. Note that $i_0+i_h\equiv 0~(\text{mod}~\lambda)$, thus, $i_0=\lambda c-b$, where $c\geq 1$. Notice that $i_h-i_0=\lambda(a-c)+2b>0$. We claim $a-c\geq -1$. Otherwise, $i_h-i_0\leq 2(b-\lambda)<0$. We continue our discussions by distinguishing the following two subcases.
      	\begin{itemize}
      		\item If $a-c\geq 0$, i.e., $1\leq c\leq a$, then $i\in \Delta_1$.
      		\item If $a-c=-1$, then $i_h-i_0=2b-\lambda>0$. It gives $\frac{\lambda}2<b<\lambda$. Hence, $i=aq^h+\frac{b(q^h-1)}\lambda+a+1$, where $0\leq a<\frac{q-1}\lambda$ and $\frac{\lambda}2<b<\lambda$. That is, $i\in \Delta_2$. 
      	\end{itemize}
      \end{itemize}
      
      The result follows.
\end{IEEEproof}\

 Note that
\begin{equation}\label{EQ1}
\begin{split}
\gcd(q-1, q^h+1)=\begin{cases}1 &\text{if} ~q ~\text{is even},\\
2 &\text{if}~q~\text{is odd}.
\end{cases}
\end{split}
\end{equation}
Define $\Delta=\left\lbrace c(q^h+1): 1\leq c\leq \frac{q-1}\lambda\right\rbrace$ when $\lambda$ is odd. Otherwise,
\[\begin{split}
\Delta=\left\lbrace \frac{c(q^h+1)}2: 1\leq c\leq \frac{2(q-1)}\lambda\right\rbrace.
\end{split}\]
We have the following conclusion.
\begin{lemma}\label{l7}
  Let $\Delta$ be defined as above. Let $m=2h\geq 4$, and $i$ be an integer with $1\leq i\leq \frac{q^{h+1}-1}\lambda$. Then $|\bC_{i}|=h$ if and only if $i\in
  \Delta $.
\end{lemma}

\begin{IEEEproof}
    Clearly, $|\bC_i|$ divides $m$. Notice that $iq^\ell<n$ for each $1\leq \ell\leq h-1$. Hence, $|\bC_i|=h$ if and only if
    \begin{equation}\label{EQ2}
    \begin{split}
    iq^h\equiv i~(\text{mod}~n) \Longleftrightarrow \lambda i\equiv 0~(\text{mod}~q^h+1).
    \end{split}
    \end{equation}
    
    \begin{itemize}
        \item [(\romannumeral1)] If $\lambda$ is odd, from (\ref{EQ1}), $\gcd(\lambda,q^h+1)=1$. It follows from (\ref{EQ2}) that
        \[\begin{split}
        iq^h\equiv i~(\text{mod}~n) \Longleftrightarrow i\equiv 0~(\text{mod}~q^h+1).
        \end{split} \]
        
        \item [(\romannumeral2)] If $\lambda$ is even, from (\ref{EQ1}), $\gcd(\lambda,q^h+1)=2$. It follows from (\ref{EQ2}) that
        \[\begin{split}
        iq^h\equiv i~(\text{mod}~n) \Longleftrightarrow i\equiv 0~(\text{mod}~(q^h+1)/2).
        \end{split}\]
    \end{itemize}
    The result follows.
\end{IEEEproof}\

Combining with Lemmas \ref{l6} and \ref{l7}, we have the following conclusion.

\begin{theorem}\label{th2}
   Let $m=2h\geq 4$. Let $i$ be an integer with $1\leq i\leq \frac{q^{h+1}-1}{\lambda}$ and $i\not\equiv 0 ~({\rm mod}~q)$. Then $i$ is not a $q$-cyclotomic coset leader modulo $n$ if and only if $i\in\Delta_0\bigcup \Delta_1\bigcup \Delta_2$, where $\Delta_0$, $\Delta_1$ and $\Delta_2$ are defined as Lemma \ref{l6}. Moreover,
    \[\begin{split}
    |\mathbb{C}_i|=\begin{cases}
    h&~{\rm if}~i\in \Delta,\\
    m~&~{\rm otherwise},
    \end{cases}
    \end{split}\]
     where $\Delta$ is defined as above. 
\end{theorem}

\begin{IEEEproof}
   The first statement of this theorem comes from Lemma \ref{l6}. We now prove that all cosets have only two possible sizes. For every integer $i$ with $1\leq i\leq \frac{q^{h+1}-1}\lambda$, it is clear that $iq^\ell <n $ for all $1\leq \ell \leq h-1$. This gives that $|\mathbb{C}_i|\geq h$. Notice that $|\mathbb{C}_i|$ divides $m$, we have $|\mathbb{C}_i|=m$ or $h$. According to Lemma \ref{l7}, the result follows.
\end{IEEEproof}\

The following corollary can be deduced from Theorem \ref{th2}.

\begin{corollary}
   Let $m=2h\geq 4$, then
    \begin{itemize}
        \item [(\romannumeral1)] if $\lambda \geq 3$ is odd, the smallest $i$ with $i\not \equiv0~({\rm mod}~q)$ that is not a $q$-cyclotomic coset leader modulo $n$ is $\frac{(\lambda+1)q^h+\lambda-1}{2\lambda}$; 
        \item [ (\romannumeral2)] if $\lambda=2$, the smallest $i$ with $i\not \equiv0~({\rm mod}~q)$ that is not a $q$-cyclotomic coset leader modulo $n$ is $\frac{3q^h+1}2$;
        \item [(\romannumeral3)] if $\lambda \geq 4$ is even, the smallest $i$ with $i\not\equiv0~({\rm mod}~q)$ that is not a $q$-cyclotomic coset leader modulo $n$ is $\frac{(\lambda+2)q^h+\lambda-2}{2\lambda}$.
    \end{itemize}
\end{corollary}

\begin{IEEEproof}
   Recall $\Delta_i$ defined as Lemma \ref{l6}. Let $\min(\Delta_i)$ denote the smallest number in $\Delta_i$ for $i=0,1,2$. It is easy to check that $f(a,b,c)$ has the following properties. If $a>a'$, then $f(a,b,c)>f(a',b',c')$ for all $0\leq b,b'<\lambda$ and $1\leq c,c'\leq q-1$. If $b>b'$, then $f(a,b,c)>f(a,b',c')$ for all $1\leq c,c'\leq q-1$. Hence, if $\Delta_i\neq \emptyset$, we have $\min(\Delta_0)=f(2,0,1)$, $\min(\Delta_1)=f(1,1,1)$ and $\min(\Delta_2)=f(0,\lceil \frac{\lambda+1}2 \rceil,1)$. This gives that the smallest $i$ with $i\not \equiv0~({\rm mod}~q)$ that is not a coset leader is $\min(\Delta_2)$ if $\lambda\geq 3$. Hence, the results of Cases  (\romannumeral1) and (\romannumeral3) are follow. 
   
   If $\lambda=2$ and $q>3$, then $\Delta_1\neq \emptyset$ and $\Delta_2=\emptyset$. It follows that the smallest $i$ with $i\not \equiv0~({\rm mod}~q)$ that is not a coset leader is $\min(\Delta_1)$. If $\lambda=2$ and $q=3$, then $\Delta_0=\Delta_1=\Delta_2=\emptyset$. That is, every integer $i\leq \frac{3^{h+1}-1}2$ with $i\not\equiv 0~(\text{mod}~3)$ is a coset leader. Notice that $\frac{(3^{h+1}+1)3^{h-1}}2\equiv \frac{3^{h-1}+1}{2} ~(\text{mod}~n)$, we have that $\frac{3^{h+1}+1}2$ is not a coset leader. The proof is completed. 
\end{IEEEproof}\

For $\lambda=1$, the result that the smallest $i$ with $i\not \equiv 0~(\text{mod}~q)$ that is not a a $q$-cyclotomic coset leader modulo $n$ is $2q^h+1$ was shown in \cite{yue1992structure}. Moreover, for $\delta\leq q^{h+1}-1$, the dimension of $\cC_{(q,m,\lambda,\delta)}$ was determined in
\cite{liu2017dimensions}. For $\lambda=q-1$, if $\delta\leq q^h$, the dimension of $\cC_{(q,m,\lambda,\delta)}$ was determine in \cite{li2017lcd}. Theorem \ref{th3} is a generalization of the results in \cite{li2017lcd}. With the conclusions on cyclotomic cosets in Theorem \ref{th2}, we determine the dimension of $\cC_{(q,m,\lambda,\delta)}$ with $\lambda\geq 2$ as follows.

\begin{theorem}\label{th3}
    Let $m=2h\geq 4$. For every integer $\delta$ with $1\leq \delta-1\leq \frac{q^{h+1}-1}\lambda$, let $\delta-1=\sum_{j=0}^h\delta_jq^j$ and $\overline{\delta}=\left\lceil \frac{(\delta-1)(q-1)}{q} \right\rceil $, where $0\leq \delta_j\leq q-1$. Then $\cC_{(q,m,\lambda,\delta)}$ has length $n=\frac{q^m-1}\lambda$, minimum distance $d\geq \delta$ and dimension $k$, where
    \begin{itemize}
        \item [(\romannumeral1)] if $\delta\leq q^h+1$, define $\varepsilon=\left\lfloor\frac{(\delta-2)\lambda}{q^h-1}\right\rfloor $, then
        \[\begin{split}
        k=\begin{cases}n-m\overline{\delta}~&{\rm if}~\varepsilon< \left\lfloor\frac{\lambda} 2\right\rfloor,\\
        n-m\overline{\delta}+m(\varepsilon-\frac{\lambda-1}{2})~&{\rm if}~\left\lfloor\frac{\lambda} 2\right\rfloor\leq \varepsilon<\lambda,\\
        n-m\overline{\delta}+m(\frac{\lambda-1}{2})~&{\rm if}~\varepsilon=\lambda.\\
        \end{cases}
        \end{split}\]
        \item [(\romannumeral2)] if $\delta\geq q^h+2$ and $\delta_h<\frac{q-1}\lambda$, define $\vartheta=\left\lfloor \frac{(\delta-2-\delta_hq^h)\lambda}{q^h-1}\right\rfloor$, then
        \[\begin{split}
       k=\begin{cases}n-m\overline{\delta}+m\frac{\lambda \delta_h^2+2(\delta_0-\delta_h)+1}{2}~{\rm if}~\delta\leq \delta_hq^h+\delta_h,\\
       n-m\overline{\delta}+m\frac{\lambda \delta_h^2}{2}~{\rm if}~\delta_hq^h+\delta_h<\delta\leq \delta_hq^h+\frac{q^h-1}\lambda+1,\\
        n-m\overline{\delta}+m\frac{\lambda \delta_h^2}{2}+m[(\vartheta-1)\delta_h+\delta_0-\frac{\vartheta (q-1)}\lambda]\\
        ~{\rm if}~\delta_hq^h+\frac{\vartheta (q^h-1)}\lambda+1<\delta\leq \delta_hq^h+\frac{\vartheta(q^h-1)}\lambda+\delta_h+1~{\rm and}~1\leq \vartheta\leq \frac{\lambda}{2},\\
        n-m\overline{\delta}+m\frac{\lambda \delta_h^2}{2}+m[(\vartheta-1)\delta_h+\delta_0-\frac{\vartheta (q-1)}\lambda]+m(\vartheta-\frac{\lambda+1}2)\\
        ~{\rm if}~\delta_hq^h+\frac{\vartheta (q^h-1)}\lambda+1<\delta\leq \delta_hq^h+\frac{\vartheta(q^h-1)}\lambda+\delta_h+1~{\rm and}~\vartheta>\frac{\lambda}{2},\\
       n-m\overline{\delta}+m\frac{\lambda \delta_h^2}{2}+m\vartheta\delta_h\\
        ~{\rm if}~\delta_hq^h+\frac{\vartheta(q^h-1)}\lambda+\delta_h+1<\delta ~{\rm and}~1\leq \vartheta<\frac{\lambda}{2},\\
        n-m\overline{\delta}+m\frac{\lambda \delta_h^2}{2}+m\vartheta\delta_h+m(\vartheta-\frac{\lambda-1}{2})\\
        ~{\rm if}~\delta_hq^h+\frac{\vartheta(q^h-1)}\lambda+\delta_h+1<\delta ~{\rm and}~\vartheta\geq \frac{\lambda}{2}.\\
        \end{cases}
        \end{split}\]
        \item [(\romannumeral3)] if $\delta\geq q^h+2$ and $\delta_h=\frac{q-1}\lambda$, then
        \[\begin{split}
        k=\begin{cases}n-m\overline{\delta}+m\frac{\lambda \delta_h^2+2(\delta_0-\delta_h)+1}{2}~&{\rm if}~\delta\leq \delta_hq^h+\delta_h,\\
        n-m\overline{\delta}+m\frac{\lambda \delta_h^2}{2}~&{\rm if}~\delta>\delta_hq^h+\delta_h.\\
        \end{cases}
        \end{split}\]
    \end{itemize}
\end{theorem}

\begin{IEEEproof}
    The lower bound on the minimum distance comes from Lemma \ref{l1}. For every integer $\delta$ with $1\leq \delta-1 \leq \frac{q^{h+1}-1}\lambda$, it follows from Theorem \ref{th2} that $|\bC_i|=m$ except for $i\in \Delta$, and $i\in \Gamma_1$ is a coset leader except for $i\in \Delta_0\bigcup\Delta_1\bigcup\Delta_2$. Hence, the dimension of the BCH code $\cC_{(q,m,\lambda,\delta)}$ is
    \begin{equation}\label{EQ3}
    \begin{split}
    k=n\!-\!m|\Gamma_1|\!+\!\frac{m}2|\Gamma_1\bigcap \Delta|\!+\!m\sum_{j=0}^2|\Gamma_1\bigcap \Delta_j|,
    \end{split}
    \end{equation}
    where $\Gamma_1$, $\Delta$, $\Delta_0$, $\Delta_1$ and $\Delta_2$ are defined as above. It is easy to see that $|\Gamma_1|=\overline{\delta}$. To determine the dimension, we just need to calculate the values of $|\Gamma_1\bigcap \Delta|$ and $|\Gamma_1\bigcap\Delta_j|$ for $j=0,1,2$, respectively.
    
    We prove the conclusion on the dimension only for the case that $\delta_hq^h+\frac{\vartheta(q^h-1)}\lambda+1<\delta\leq \delta_hq^h+\frac{\vartheta(q^h-1)}\lambda+\delta_h+1$ and $\vartheta>\frac{\lambda}{2}$, where $\delta_h<\frac{q-1}\lambda$, $\lambda\geq 2$ is even integer. The proofs of the other cases are similar, and details are omitted here. 
  
    Let $\Gamma=\{i: 1\leq i\leq \delta_h q^h, \text{ and } i\not\equiv 0~(\text{mod}~q)\}$ and $\Gamma'=\{i: \delta_hq^h+1\leq i\leq \delta-1, \text{ and}~ i\not\equiv 0~(\text{mod}~q)\}$, then $\Gamma_1=\Gamma \bigcup \Gamma'$. It follows from (\ref{EQ3}) that
    \[\begin{split}
    k=n-m\overline{\delta}+\frac{m}2|\Gamma\bigcap \Delta|+m\sum_{j=0}^2|\Gamma\bigcap \Delta_j|\\
    +\frac{m}2|\Gamma'\bigcap\Delta|+m\sum_{j=0}^2|\Gamma'\bigcap \Delta_j|.
    \end{split}\]
    It is easy to check the following results are established.
    \[\begin{split}
    \Gamma\bigcap \Delta=\left\lbrace \frac{c(q^h+1)}{2}:1\leq c\leq 2\delta_h-1\right\rbrace,
    \end{split}\]
    \[\begin{split}
     \Gamma\bigcap\Delta_0=\left\lbrace f(a,0,c):1\leq c<a\leq \delta_h-1\right\rbrace,
     \end{split}\]
    \[\begin{split}
    \Gamma\bigcap\Delta_1=\left\lbrace f(a,b,c):1\leq c\leq a\leq\delta_h-1,1\leq b<\lambda\right\rbrace,
    \end{split}\]
    and
    \[\begin{split}
    \Gamma\bigcap \Delta_2=\left\lbrace f(a,b,a+1):0\leq a\leq \delta_h-1,\frac{\lambda}2<b<\lambda\right\rbrace,
    \end{split}\] 
    where $f(a,b,c)$ is defined as Lemma \ref{l6}. It follows that
   \[\begin{split}
   k=&n-m\overline{\delta}+m\left(  \frac{\lambda
   	\delta_h^2-2\delta_h+1}{2}\right)\\
   &+\frac{m}2\left| \Gamma'\bigcap\Delta \right|
   +m\sum_{i=0}^2\left| \Gamma'\bigcap \Delta_i \right|.
   \end{split}\]
    
     For $\delta_hq^h+\frac{\vartheta(q^h-1)}\lambda+1<\delta\leq \delta_hq^h+\frac{\vartheta(q^h-1)}\lambda+\delta_h+1$ and $\vartheta>\frac{\lambda}{2},$ we have 
     \[\begin{split}
    \delta-1=\delta_hq^h+\frac{\vartheta (q-1)}\lambda[q^{h-1}+\cdots+q]+\delta_0,
     \end{split}\] 
     where $\frac{\vartheta(q-1)}\lambda+1\leq \delta_0\leq \frac{\vartheta(q-1)}\lambda+\delta_h$. Clearly,
     \[\begin{split}
     \Gamma'\bigcap\Delta=\left\lbrace \delta_h(q^h+1), \frac{(2\delta_h+1)(q^h+1)}2 \right\rbrace,
     \end{split}\] 
     \[\begin{split}
     \Gamma'\bigcap \Delta_0=\left\lbrace f(\delta_h,0,c):1\leq c<\delta_h\right\rbrace,
     \end{split}\] 
     \[\begin{split}
     \Gamma'\bigcap \Delta_1=\left\lbrace f(\delta_h,b,c):1\leq c\leq \delta_h,1\leq b\leq \vartheta-1\right\rbrace\\
     \bigcup \left\lbrace f(\delta_h,\vartheta,c):1\leq c\leq \delta_0-\frac{\vartheta(q-1)}{\lambda}\right\rbrace.
     \end{split}\]
     and
     \[\begin{split}
     \Gamma'\bigcap\Delta_2=\left\lbrace f(\delta_h,b,\delta_h+1): \frac{\lambda+2}2\leq b\leq \vartheta-1\right\rbrace.
     \end{split}\]
     It follows that 
     \[\begin{split}
     k=n-m\overline{\delta}+m\frac{\lambda \delta_h^2}{2}+m\left[ (\vartheta-1)\delta_h+\delta_0-\frac{\vartheta (q-1)}\lambda\right]+m\left( \vartheta-\frac{\lambda+1}2\right). 
    \end{split}\] 
     The result follows.
\end{IEEEproof}\

\section{The weight distribution of two classes of BCH codes} \label{sec4}

In this section, we study the weight distribution of BCH codes of length $n=\frac{q^m-1}\lambda$, where $m\geq 2$ is an integer. Our main task is to find a trace representation for the codewords in this class of BCH codes. For this reason, we need to find the first few largest $q$-cyclotomic coset leaders modulo $n$. 

When $\lambda=1$, the first few largest $q$-cyclotomic coset leaders modulo $n$ were determined in \cite{ding2017dimension} and \cite{li2017minimum}. When $\lambda=2$ and $q=3$, the first few largest $q$-cyclotomic coset leaders modulo $n$ were determined in \cite{li2016narrow}. It seems to be a hard problem to determine the first few largest $q$-cyclotomic coset leaders modulo $n$ for all $q$, $m$ and $\lambda$. We only deal with the cases $\lambda=2$ and $\lambda=q-1$.

 For every integer $i$ with $0\leq i\leq n-1$, the $q$-adic expansion of $i$ is defined by $\sum_{\ell=0}^{m-1}i_{\ell}q^{\ell}$, where $0\leq i_{\ell}\leq q-1$. We will study the properties of the cyclotomic cosets by using $q$-adic expansion in the following paper. Let $[a]_n$ be the smallest non-negative integer such that $a\!\equiv\! [a]_n~(\text{mod}~n)$.

\begin{lemma}\label{l8}
	Let $1\leq i\leq n-1$ be an integer. Denote the $q$-adic expansion of $i$ by $\sum_{\ell=0}^{m-1}i_{\ell}q^{\ell}$. If $i$ is a $q$-cyclotomic coset leader modulo $n$, then $0\leq i_{m-1}\leq \frac{q-1}\lambda-1$ and $i_{\ell}\geq i_{m-1}$ for all $0\leq \ell\leq m-2$. 
\end{lemma}

\begin{IEEEproof}
	Note that $n\!=\!(\frac{q-1}\lambda) \sum_{\ell=0}^{m-1}q^{\ell}$. From $i\!\leq\! n-1$, there exists an index $v$ with $0\leq v\leq m-1$ such that $i_{v}\leq \frac{q-1}\lambda-1$. If $v=m-1$, then $i_{m-1}\leq \frac{q-1}\lambda-1$. If $v\leq m-2$, we have 
	\[ [iq^{m-1-v}]_n=\sum_{\ell=0}^{v}i_{\ell}q^{m-1-v+\ell}+\sum_{\ell=v+1}^{m-1}i_{\ell}q^{\ell-v-1}.\]
    From $i\leq [iq^{m-1-v}]_n$, we deduce $i_{m-1}\leq i_{v}\leq \frac{q-1}\lambda-1$. 
    
    We now prove $i_{\ell}\geq i_{m-1}$ for all $0\leq \ell \leq m-2$. If there is an index $u$ such that $i_{u}<i_{m-1}$, then $[iq^{m-1-u}]_{n}<i$, which contradicts the fact that $i$ is a coset leader.
\end{IEEEproof}	

\subsection{The weight distribution of BCH codes of length $(q^m-1)/2$}

Throughout this subsection, let $q$ be an odd prime power and $m\geq 2$ be an integer. We will find the first few largest $q$-cyclotomic coset leaders modulo $n=\frac{q^m-1}2$. Let $\delta_i$ denote the $i$-th largest coset leader, then $\delta_1$, $\delta_2$ and $\delta_3$ are explicitly given in \cite{li2016narrow} when $q=3$.

\begin{lemma}\label{l9}
    The largest $q$-cyclotomic coset leader modulo $n=\frac{q^{m}-1}2$ is 
    \[\begin{split}
    \delta_1=\frac{q^m-1-q^{m-1}-q^{\lfloor \frac{m-1}2\rfloor}}2.
    \end{split}\]
    Furthermore, $\left| \bC_{\delta_1}\right|=m $ when $m$ is odd and $\left|\bC_{\delta_1}\right|=\frac{m}2 $ when $m$ is even.
\end{lemma}

\begin{IEEEproof}
    When $q=3$, $\delta_1$ was determined in \cite{li2016narrow}. Now assume $q\geq 5$, we distinguish two cases for even and odd $m$.
  
  Case 1. $m\geq 3$ is odd. It is easy to see that
   \[\begin{split}
   \bC_{\delta_1}=\left\lbrace \frac{q^m-1-q^{\ell-1}-q^{\ell+\frac{m-1}2}}2:
   1\leq \ell \leq \frac{m-1}2\right\rbrace\\
   \bigcup \left\lbrace \frac{q^m-1-q^{\ell-1}-q^{\ell-\frac{m+1}2}}2:
   \frac{m+1}2\leq \ell \leq m\right\rbrace,
   \end{split}\]
    and the $q$-adic expansion of $\delta_1$ is 
    \[\begin{split}
    \left( \frac{q-3}2\right) q^{m-1}+(q-1)\sum_{i=\frac{m-1}2}^{m-2}q^i+\left( \frac{q-1}2\right) \sum_{i=0}^{\frac{m-3}2}q^i.
    \end{split}\]
    Hence, $|\bC_{\delta_1}|=m$ and $\delta_1$ is the smallest integer in $\bC_{\delta_1}$. We will prove that $\delta_1$ is the largest integer in the set of all coset leaders. Suppose there exists an integer $s$ with $\delta_1\!<\!s\!<\!n$ is a $q$-cyclotomic coset leader modulo $n$, by Lemma \ref{l8}, the $q$-adic expansion of $s$ must be of the form
    \[\begin{split}
    \left( \frac{q-3}2\right) q^{m-1}+(q-1)\sum_{i=\frac{m-1}2}^{m-2}q^i+ \sum_{i=0}^{\frac{m-3}2}s_iq^i,
    \end{split}\]
    where $s_i\geq \frac{q-3}2$ and $\sum_{i=0}^{\frac{m-3}2}s_iq^i>(\frac{q-1}2)\sum_{i=0}^{\frac{m-3}2}q^i$. 
    
    When $m=3$, we have $\frac{q+1}2\leq s_0\leq (q-1)$. Moreover,
    \[\begin{split}
    [sq^2]_n=\left( s_0-\frac{q+1}2\right) q^2+(q-1)q+\frac{q-1}2.
    \end{split}\]
    It follows that $[sq^2]_n\leq \delta_1<s$, and so we arrive at a contradiction. Now consider the case $m\geq 5$ in the following.
    
     Case 1.1. There exists an index $v$ such that $s_{v}=\frac{q-3}2$. From $s>\delta_1$, we obtain $0\leq v\leq \frac{m-5}2$, and
    \[\begin{split}
    [sq^{m-1-v}]_n=\sum_{i=0}^{v} s_iq^{i+m-1-v}+\left( \frac{q-3}2\right) q^{m-2-v}\\
    +(q-1)\sum_{i=\frac{m-1}2}^{m-2}q^{i-1-v}+\sum_{i=v+1}^{\frac{m-3}2}s_iq^{i-v-1}.
    \end{split}\]
    Note that $m-2-v\geq \frac{m+1}2$, we have $[sq^{m-1-v}]_n<s$, which gives a contradiction.
    
    Case 1.2. There exists an index $v$ with $0\leq v \leq \frac{m-3}2$ such that $\frac{q-1}2<s_{v}<q-1$. Then
    \[\begin{split}
    &[sq^{m-1-v}]_n=\sum_{i=0}^v \left( s_i-\frac{q-1}2\right) q^{i+m-1-v}-q^{m-2-v}\\
    &+\left( \frac{q-1}2\right)\sum_{i=\frac{m-1}2}^{m-2}q^{i-1-v}
    +\sum_{i=v+1}^{\frac{m-3}2}\left( s_{i}-\frac{q-1}2\right) q^{i-1-v}\\
   & \leq \left( \frac{q-3}2\right)q^{m-1}+\left( \frac{q-1}2\right) \sum_{i=0}^{m-2} q^i<\delta_1.
    \end{split}\]  
   Hence, $[sq^{m-1-v}]_n<s$, a contradiction. 
    
   Case 1.3. There exists an index $v$ with $1\leq v\leq \frac{m-3}2$ such that $s_v=\frac{q-1}2$ and $s_{v-1}=q-1$. Similar to Case 1.2, we have $[sq^{m-1-\nu}]_n<s$, a contradiction.  
   
    Summarizing the discussions above, we just need to prove that for $s=(\frac{q-3}2)q^{m-1}+(q-1)\sum_{i=v}^{m-2}q^i+(\frac{q-1}2)\sum_{i=0}^{v-1}q^i$, there exists an integer $i\in \bC_s$ such that $i<s$, where $0\leq v\leq \frac{m-5}2$. At this point,
   \[\begin{split}
   [sq^{m-1-v}]_n=\frac{q^m-1-q^{m-1}-q^{m-2-v}}2\\
   \leq \frac{q^m-1-q^{m-1}-q^{\frac{m+1}2}}2<\delta_1<s.
   \end{split}\]
    This gives a contradiction.
    
    Collecting all the conclusions above, we conclude that $\delta_1$ is the largest coset leader for the case that $m$ is odd.
 
    Case 2. $m\geq 2$ is even. It is easy to see that
    \[\begin{split}
    \bC_{\delta_1}=\left\lbrace \frac{q^m-1-q^{\ell-1}-q^{\ell+\frac{m}2-1}}2: 1\leq \ell \leq \frac{m}2\right\rbrace.
    \end{split}\]
    Clearly, $|\bC_{\delta_1}|=\frac{m}2$ and $\delta_1$ is the coset leader of $\bC_{\delta_1}$. Similarly as in the case that $m$ is odd, one can prove that $\delta_1$ is the largest coset leader for the case that $m$ is even. Details are omitted here.
    
This completes the proof.
\end{IEEEproof}\

  Similarly, we can calculate the second and the third largest $q$-cyclotomic coset leaders modulo $\frac{q^m-1}2$. 

\begin{lemma}\label{l10}
    The second largest $q$-cyclotomic coset leader modulo $n=\frac{q^m-1}2$ is
    \[\begin{split}
    \delta_2=\frac{q^{m}-1-q^{m-1}-q^{\lfloor \frac{m+1}2\rfloor}}2
    \end{split}\] 
    and $|\bC_{\delta_2}|=m$.
\end{lemma}

\begin{IEEEproof}
    When $q=3$, $\delta_2$ was determined in \cite{li2016narrow}. Now consider the case $q\geq 5$. The proof is divided into the following two cases according to the parity of $m$.
   
    Case 1. $m\geq 3$ is odd. It is easy to see that
    \[\begin{split}
    \bC_{\delta_2}=\left\lbrace \frac{q^m-1-q^{\ell-1}-q^{\ell+\frac{m+1}2}}2: 1\leq \ell \leq \frac{m-3}2\right\rbrace\\
    \bigcup \left\lbrace \frac{q^m-1-q^{\ell-1}-q^{\ell-\frac{m-1}2}}2: \frac{m-1}2\leq \ell \leq m\right\rbrace,
    \end{split}\]
    and the $q$-adic expansion of $\delta_2$ is
    \[\begin{split}
    \left( \frac{q-3}2\right)q^{m-1}+(q-1) \sum_{i=\frac{m+1}2}^{m-2}q^i+\left( \frac{q-1}2\right) \sum_{i=0}^{\frac{m-1}2}q^i.
    \end{split}\]
    Therefore,  $|\bC_{\delta_2}|=m$ and $\delta_2$ is the smallest integer in $\bC_{\delta_2}$. Suppose there exists an integer $s$ with $\delta_2<s<\delta_1$ is a $q$-cyclotomic coset leader modulo $n$, by Lemma \ref{l8}, the $q$-adic expansion of $s$ must be of the form
    \[\begin{split}
    \left( \frac{q-3}2\right) q^{m-1}+\left( q-1\right) \sum_{i=\frac{m+1}2}^{m-2}q^i+\sum_{i=0}^{\frac{m-1}2}s_iq^i,
    \end{split}\]
    where $\frac{q-3}2\leq s_i\leq q-1$ and
    \begin{equation}\label{EQ4}
    \begin{split}
    \left( \frac{q-1}2\right) \sum_{i=0}^{\frac{m-1}2}q^i<\sum_{i=0}^{\frac{m-1}2}s_iq^i<\left( q-1\right)q^{\frac{m-1}2}+\left( \frac{q-1}2\right) \sum_{i=0}^{\frac{m-3}2}q^i.
    \end{split}
    \end{equation}
    
     We continue our discussions by distinguishing the following three cases.
     
     Case 1.1. $s_{\frac{m-1}2}=\frac{q-1}2$. From (\ref{EQ4}), there exists an index $\iota$ with $0\leq \iota \leq \frac{m-3}2$ such that $s_\iota >\frac{q-1}2$. Let $v$ be the largest index such that $s_{v}>\frac{q-1}2$. Similar to the proof of Case 1.2 in Lemma \ref{l9}, we have $[sq^{m-2-v}]_n<s$, which contradicts the fact that $s$ is a coset leader.
    
     Case 1.2. $\frac{q-1}2<s_{\frac{m-1}2}<q-1$. Similar to Case 1.2 in Lemma \ref{l9}, we have $[sq^{\frac{m-1}2}]_n<s$, a contradiction. 
    
     Case 1.3. $s_{\frac{m-1}2}=q-1$. From (\ref{EQ4}), there exists an index $\iota$ with $0\leq \iota \leq \frac{m-3}2$ such that $s_\iota<\frac{q-1}2$. Let $v$ be the largest index such that $s_{v}<\frac{q-1}2$, Similar to Lemma \ref{l9}, one can prove that $[sq^{m-1-v}]_n<s$. Hence, $s$ cannot be a coset leader.
   
    Summarizing all the conclusion above, we obtain that $\delta_2$ is the second largest coset leader for the case that $m$ is odd.

   Case 2. $m\geq 2$ is even. It is easy to see that
   \[\begin{split}
   \bC_{\delta_2}=\left\lbrace \frac{q^m-1-q^{\ell-1}-q^{\ell+\frac{m}2}}2: 1\leq \ell \leq \frac{m-2}2\right\rbrace \\
   \bigcup \left\lbrace\frac{q^m-1-q^{\ell-1}-q^{\ell-\frac{m}2}}2: \frac{m}2\leq \ell \leq m\right\rbrace.
   \end{split}\]
    Hence, $|\bC_{\delta_2}|=m$ and $\delta_2$ is the coset leader in $\bC_{\delta_2}$. Similarly as in the case that $m$ is odd, one can prove that $\delta_2$ is the second largest coset leader for the case that $m$ is even. Details are omitted here.
 
   The desired result follows.
\end{IEEEproof}\


\begin{lemma}\label{l11}
    Let $m\geq 6$. Then the third largest $q$-cyclotomic coset leader modulo $n=\frac{q^m-1}2$ is 
    \[\begin{split}
    \delta_3=\frac{q^{m}-1-q^{m-1}-q^{\lfloor \frac{m+3}2\rfloor}}2.
    \end{split}\]
    In addition, $|\bC_{\delta_3}|=m$.
\end{lemma}

\begin{IEEEproof}
    When $q=3$, $\delta_3$ was determined in \cite{li2016narrow} for $m\geq 9$. We
    can verify that $\delta_3$ is the third largrst coset leader for the case that $6\leq m\leq 8$. Now consider the case $q\geq 5$. The proof is divided into the following two cases.

    Case 1. $m\geq 7$ is odd. We have
    \[\begin{split}
    \bC_{\delta_3}=\left\lbrace \frac{q^m-1-q^{\ell-1}-q^{\ell+\frac{m+3}2}}2:1\leq \ell \leq \frac{m-5}2\right\rbrace\\
    \bigcup\left\lbrace \frac{q^m-1-q^{\ell-1}-q^{\ell-\frac{m-3}2}}2:\frac{m-3}2\leq \ell \leq m\right\rbrace
    \end{split}\]
    and the $q$-adic expansion of $\delta_3$ is
    \[\begin{split}
    \left(\frac{q-3}2\right) q^{m-1}+(q-1) \sum_{i=\frac{m+3}2}^{m-2}q^i+\left( \frac{q-1}2\right)\sum_{i=0}^{\frac{m+1}2}q^i.
    \end{split}\]
    Therefore, $|\bC_{\delta_3}|=m$ and $\delta_3$ is the coset leader in $\bC_{\delta_3}$. Suppose there exists an integer $s$ with $\delta_3<s<\delta_2$ is a $q$-cyclotomic coset leader modulo $n$, by Lemma \ref{l8}, the $q$-adic expansion of $s$ must be of the form
    \[\begin{split}
    \left( \frac{q-3}2\right) q^{m-1}+\left( q-1\right) \sum_{i=\frac{m+3}2}^{m-2}q^i+\sum_{i=0}^{\frac{m+1}2}s_iq^i,
    \end{split}\]
    where $\frac{q-3}2\leq s_i\leq q-1$ and
    \begin{equation}\label{EQ5}
    \begin{split}
    \left( \frac{q-1}2\right) \sum_{i=0}^{\frac{m+1}2}q^i<\sum_{i=0}^{\frac{m+1}2}s_iq^i<\left( q-1\right)q^{\frac{m+1}2}+\left( \frac{q-1}2\right) \sum_{i=0}^{\frac{m-1}2}q^i.
    \end{split}
    \end{equation}
   
    Similar to Lemma \ref{l10}, we can prove that the following Cases 1.1 and 1.2 are hold. 
 
    Case 1.1. $s_{\frac{m+1}2}=\frac{q-1}2$. Let $v$ be the largest index such that $s_{v}>\frac{q-1}2$, then $[sq^{m-2-v}]_n<s$, which gives a contradiction.
    
    Case 1.2. $\frac{q-1}2<s_{\frac{m+1}2}<q-1$. Then $[sq^{\frac{m-3}2}]_n<s$, we obtains a contradiction.
    
    Case 1.3. $s_{\frac{m+1}2}=q-1$. From (\ref{EQ5}), there exists an index $\iota$ with $0\leq \iota \leq \frac{m-1}2$ such that $s_\iota=\frac{q-3}2$. Let $v$ be the largest index such that $s_{v}=\frac{q-3}2$. It follows that $s_i=\frac{q-1}2$ for all $v+1\leq i\leq \frac{m-1}2$. Then,
   \[\begin{split}
   [sq^{m-1-v}]_n=\sum_{i=0}^{v}s_iq^{m-1-v+i}+\left( \frac{q-3}2\right)q^{m-2-v}\\
   +\left( q-1\right) \sum_{i=\frac{m+1}2}^{m-2}q^{i-1-v}+\left( \frac{q-1}2\right) \sum_{i=v+1}^{\frac{m-1}2}q^{i-1-v}.
   \end{split}\]
    If $v\leq \frac{m-3}2$, we deduce $[sq^{m-1-v}]_n<s$ since $m-2-v\geq v+1$. If $v=\frac{m-1}2$, we continue our discussions of this case by distinguishing the following cases.
    
    Case 1.3.1. If there exists an index $v$ with $1\leq v\leq \frac{m-3}2$ such that $s_v<q-1$, we have $[sq^{\frac{m-1}2}]_n<s$, a contradiction.
    
    Case 1.3.2. If $s_{\frac{m-1}2}=\frac{q-3}2$ and $s_i=q-1$ for all $1\leq i\leq \frac{m-3}2$, we have $[sq^{\frac{m-3}2}]_n<s$ since $m-2>\frac{m+1}2$.
   
    Summarizing all the conclusion above, we obtain that $\delta_3$ is the third largest coset leader for the case that $m$ is odd.

     Case 2. $m\geq 6$ is even. It is easy to see that
    \[\begin{split}
    \bC_{\delta_3}=\left\lbrace \frac{q^m-1-q^{\ell-1}-q^{\ell+\frac{m+2}2}}2: 1\leq \ell \leq \frac{m-4}2\right\rbrace\\
    \bigcup \left\lbrace \frac{q^m-1-q^{\ell-1}-q^{\ell-\frac{m-2}2}}2: \frac{m-2}2\leq \ell \leq m\right\rbrace
    \end{split}\]
    Obviously, $|\bC_{\delta_3}|=m$ and $\delta_3$ is the coset leader in $\bC_{\delta_3}$. Similarly as in the case that $m$ is odd, one can prove that $\delta_3$ is the third largest coset leader for the case that $m$ is even. Details are omitted here.
    
    The desired result follows.
\end{IEEEproof}\

Based on the lemmas above, we can calculate the weight distribution of BCH code $\cC_{(q,m,2,\delta_i)}$ and $\widehat{\cC}_{(q,m,2,\delta_i)}$ as follows.

\begin{theorem}\label{th4}
    The BCH code $\widehat{\cC}_{(q,m,2,\delta_1)}$ has parameters $[\frac{q^m-1}2,k,d]$, where
    \begin{itemize}
        \item [(\romannumeral1)] if $m$ is odd, then $k=m$ and $d=\frac{(q-1)q^{m-1}}2$.
        \item [(\romannumeral2)] if $m$ is even, then $k=\frac{m}2$ and $d=\frac{(q-1)(q^{m-1}+q^{k-1})}{2}$.
    \end{itemize}
    In addition, $\widehat{\cC}_{(q,m,2,\delta_1)}$ has only one nonzero weight, and meets the Griesmer bound.
\end{theorem}

\begin{IEEEproof}
     Let $\alpha$ be a primitive element of $\Fqm$, then $\alpha^2$ is a primitive $n$-th root of unity in $\Fqm$. From Lemma \ref{l9}, the code $\widehat{\cC}_{(q,m,2,\delta_1)}$ has one nonzero and $\alpha^{2\delta_1}$ is a root of its parity-check polynomial. The dimension of $\widehat{\cC}_{(q,m,2,\delta_1)}$ follows from Lemma \ref{l9}.
    
    Case 1. $m$ is odd. Notice that $\gcd(2\delta_1,q-1)=\gcd(2,q-1)=2$, and $\gcd(2\delta_1,\frac{q^m-1}{q-1})=\gcd(q^{\frac{m-1}2}+1,\frac{q^m-1}{q-1})=\gcd(q^{\frac{m-1}2}+1,q^{m-1})=1$. From Theorem 11 in \cite{vaga2007}, the result follows.

    Case 2. $m$ is even. Let $h=\frac{m}2$ and $\tau=q^{m-1}+q^{h-1}$, then $-2\delta_1\equiv \tau~(\text{mod } q^m-1)$. By Lemma \ref{l3},
    \[\begin{split}
    \widehat{\cC}_{(q,m,2,\delta_1)}=\left\lbrace \left(\text{Tr}^{q^h}_q(a\alpha^{\tau\ell})\right)_{\ell=0}^{n-1} : a\in \ff_{q^{h}}\right\rbrace.
    \end{split}\]
    
    Let $\beta=\alpha^{(q^h+1)}$. Since $\text{Tr}^{q^h}_q(a\alpha^{\tau\ell})=\text{Tr}^{q^h}_q(a^q\alpha^{q\tau\ell})$, it follows that $\widehat{\cC}_{(q,m,2,\delta_1)}$ has the same weight distribution with the following code
    \[\begin{split}
    \left\lbrace c(a)=\left(\text{Tr}^{q^h}_q(a\beta^{\ell})\right)_{\ell=0}^{n-1}: a\in \ff_{q^h} \right\rbrace.
    \end{split}\]
    Let $n'=q^h-1$ and 
    \[\begin{split}
    C'=\left\lbrace c'(a)=\left(\text{Tr}^{q^h}_q(a\beta^{\ell})\right)_{\ell=0}^{n'-1}: a\in \ff_{q^h} \right\rbrace.
    \end{split} \]
   Clearly,
    \[\begin{split}
   w(c'(a))&=n'-\left| \left\lbrace \ell: \text{Tr}^{q^{h}}_q(a\beta^{\ell})=0,0\leq \ell\leq n'-1 \right\rbrace  \right|\\
   &=n'-\left| \left\lbrace x\in \ff_{q^h}^*: \text{Tr}^{q^{h}}_q(ax)=0 \right\rbrace  \right|.
    \end{split}\]
    Hence, $C'$ is a $[q^h-1, h, q^h-q^{h-1}]$ one-weight code over $\mathbb{F}_q$. It is easy to check that
    \[\begin{split}
    c(a)=\overbrace{c'(a)\parallel \cdots \parallel c'(a)}^{(q^h+1)/2},
    \end{split} \]
    where $\parallel$ denotes the concatenation of vectors. Hence, $\widehat{\cC}_{(q,m,2,\delta_1)}$ is a $[n,h, \frac{(q-1)(q^{m-1}+q^{h-1})}2]$ one-weight code over $\mathbb{F}_q$.
    
    Let $C$ be a linear code of length $n$ over $\ff_q$ with dimension $k$ and minimum distance $d$. Recall the Griesmer bound (see \cite{griesmer1960}) for $C$ is $n\geq \sum_{i=0}^{k-1}\lceil \frac{d}{q^i} \rceil$, where $\lceil x \rceil$ denotes the smallest integer greater than or equal to $x$. It is easy to check that $\widehat{\cC}_{(q,m,2,\delta_1)}$ meets the Griesmer bound, and hence is optimal.
\end{IEEEproof}\

\begin{remark}
	It is well known that all one-weight code with dual weight at least $2$ have been completely characterized by Wolfmann \cite{wolfmann2005}. Moreover, a set of characterizations for the one-weight irreducible cyclic codes was introduced
	by Vega \cite{vaga2007}. Hence, the result of Theorem \ref{th4} is not new. However, we show that this class of BCH code is also one-weight code.
\end{remark}

  In the following theorem, we calculate the weight distribution of BCH code $\cC_{(q,m,2,\delta_1)}$.

\begin{theorem}\label{th5}
    The BCH code $\cC_{(q,m,2,\delta_1)}$ has parameters $[\frac{q^m-1}2,k,\delta_1]$, where
    \begin{itemize}
        \item [(\romannumeral1)] if $m$ is odd, then $k=m+1$ and $\cC_{(q,m,2,\delta_1)}$ is a four-weight code. In addition, the weight distribution of $\cC_{(q,m,2,\delta_1)}$ is listed in Table \ref{Tab2}.
        \item [(\romannumeral2)] if $m$ is even, then $k=\frac{m}2+1$ and $\cC_{(q,m,2,\delta_1)}$ is a three-weight code if $m\geq 4$. In addition, the weight distribution of $\cC_{(q,m,2,\delta_1)}$ is listed in Table \ref{Tab3}.
    \end{itemize}
\end{theorem}

\begin{IEEEproof}
     Let $\alpha$ be a primitive element of $\Fqm$, then $\alpha^2$ is a primitive $n$-th root of unity in $\Fqm$. From Lemma \ref{l9}, the code $\cC_{(q,m,2,\delta_1)}$ has two nonzeros, and $1$ and $\alpha^{2\delta_1}$ are two non-conjugate roots of its parity-check polynomial. The dimension of $\cC_{(q,m,2,\delta_1)}$ follows from Lemma \ref{l9}.
  
    Case 1. $m$ is odd. Let $\tau=q^{m-1}+q^{\frac{m-1}2}$. From Lemmas \ref{l3} and \ref{l9},
    \[\begin{split}
    \cC_{(q,m,2,\delta_1)}=\left\lbrace \left(\text{Tr}^{q^m}_q(a\alpha^{\tau\ell})+b\right)_{\ell=0}^{n-1}: a\in \Fqm, b\in \Fq \right\rbrace.
    \end{split}\] 
    Note that $\gcd(\tau,q^m-1)=\gcd(q^{\frac{m-1}2}+1,q^m-1)=\gcd(q^{\frac{m-1}2}+1,q-1)=2$. Hence, $\cC_{(q,m,2,\delta_1)}$ has the same weight distribution with the following code
    \[\begin{split}
    \left\lbrace c(a,b)=\left(\text{Tr}^{q^m}_q(a\alpha^{2\ell})+b\right)_{\ell=0}^{n-1}: a\in \Fqm, b\in \Fq \right\rbrace .
    \end{split}\]
    
    If $a=0$, then $w(c(a,b))=\frac{q^m-1}2$ for each $b\in \Fq^*$. If $b=0$, from Theorem \ref{th4}, $w(c(a,b))=\frac{(q-1)q^{m-1}}2$ for each $a\in \Fqm^*$. If $a\neq 0$ and $b\neq 0$, then
    \[\begin{split}
    w(c(a,b))&=n-\sum_{\ell=0}^{n-1}\frac{1}q\sum_{y\in \Fq}\zeta_p^{\text{Tr}_p^q( y\text{ Tr}^{q^m}_q(a\alpha^{2\ell})+yb) }\\
    &=n-\frac{1}q\sum_{y\in \Fq}\zeta_p^{\text{Tr}_p^q(yb)}\sum_{\ell=0}^{n-1}\zeta_p^{\text{Tr}^{q^m}_p( ay\alpha^{2\ell}) }\\
     &=\frac{(q-1)n}q-\frac{1}q\sum_{y\in \Fq^*}\zeta_p^{\text{Tr}_p^q(yb)}\sum_{\ell=0}^{n-1}\zeta_p^{\text{Tr}^{q^m}_p( ay\alpha^{2\ell}) }.
    \end{split}\]
    Note that
    \begin{align}\label{EQ6}
    \begin{split}
    &\sum_{x\in \ff_{q^m}^*}\zeta_p^{\text{Tr}^{q^m}_p( ayx^2) }=\sum_{\ell=0}^{2n-1}\zeta_p^{\text{Tr}^{q^m}_p( ay\alpha^{2\ell}) }\\
    =&\sum_{\ell=0}^{n-1}\left[ \zeta_p^{\text{Tr}^{q^m}_p( ay\alpha^{2\ell}) }+\zeta_p^{\text{Tr}^{q^m}_p( ay\alpha^{2(\ell+n)})}\right] 
    \end{split}
    \end{align}
    and $\text{ord}(\alpha)=2n$, thus, 
    \[\begin{split}
    \sum_{\ell=0}^{n-1} \zeta_p^{\text{Tr}^{q^m}_p( ay\alpha^{2\ell}) }=\frac{1}2\sum_{x\in \ff_{q^m}^*}\zeta_p^{\text{Tr}^{q^m}_p( ayx^2)}.
    \end{split}\]  
    It follows that 
    \[\begin{split}
     w(c(a,b))=\frac{(q-1)n}q-\frac{1}{2q}\sum_{y\in \Fq^*}\zeta_p^{\text{Tr}_p^q(yb)}\sum_{x\in \ff_{q^m}^*}\zeta_p^{\text{Tr}^{q^m}_p( ayx^2)}.
    \end{split}\]
    From Lemma \ref{l2},
     \[\begin{split}
   \sum_{x\in \ff_{q^m}^*}\zeta_p^{\text{Tr}^{q^m}_p( ayx^2)}=\eta(ay)G_{q^m}-1.
    \end{split}\]
    where $\eta$ is the quadratic character of $\Fqm$. We define a function $\eta'(y)=\eta(y)$, $y\in \Fq$. Since $m$ is odd, $\eta'$ is the quadratic character of $\Fq$. Hence,
     \[\begin{split}
    w(c(a,b))=\frac{(q-1)n}q-\frac{1}{2q}\sum_{y\in \Fq^*}\zeta_p^{\text{Tr}_p^q(yb)}\left(\eta(ay)G_{q^m}-1 \right) \\
    =\frac{(q-1)n}q-\frac{1}{2q}-\frac{\eta(a)G_{q^m}}{2q}\sum_{y\in \Fq^*}\zeta_p^{\text{Tr}_p^q(yb)}\eta(y).
    \end{split}\]
    Note that
    \[\begin{split}
    \sum_{y\in \Fq^*}\zeta_p^{\text{Tr}_p^q(yb)}\eta(y)=\sum_{y\in \Fq^*}\zeta_p^{\text{Tr}_p^q(yb)}\eta'(y)\\
    =\eta'(b^{-1})G_q=\eta(b)G_q,
    \end{split}\]
    we have $w(c(a,b))=\frac{q^m-q^{m-1}-1}2-\frac{\eta(ab)G_qG_{q^m}}{2q}$.
    
    Assume $s=[\Fq:\ff_p]$, from Lemma \ref{l2}, if $p\equiv 1~(\text{mod}~4)$, then $w(c(a,b))=\frac{q^m-q^{m-1}-\eta(ab)q^{\frac{m-1}2}-1}2$. If $p\equiv 3~(\text{mod}~4)$, then
    \[\begin{split}
    w(c(a,b))=\frac{q^m-q^{m-1}-\eta(ab)(-1)^{\frac{(m+1)s}{2}}q^{\frac{m-1}2}-1}2.
    \end{split}\]
   
    Let $n_\varepsilon=\left| \left\lbrace (a,b)\in \Fqm^*\times\Fq^*: \eta(ab)=\varepsilon \right\rbrace \right| $, where $\varepsilon=1$ or $-1$, it is easy to check that $n_1=n_{-1}=\frac{(q-1)(q^m-1)}{2}$. The weight distribution then follows.
    
    Case 2. $m$ is even. Similar to Case 1, $\cC_{(q,m,2,\delta_1)}$ has the same weight distribution with the following code
    \[\begin{split}
    \left\lbrace c(a,b)=\left(\text{Tr}^{q^h}_q(a\beta^{\ell})+b\right)_{\ell=0}^{n-1}: a\in \ff_{q^h}, b\in \Fq\right\rbrace,
    \end{split}\]
    where $h=\frac{m}2$ and $\beta=\alpha^{(q^h+1)}$.
    
    If $a=0$, we have $w(c(a,b))=\frac{q^m-1}2$ for each
    $b\in \Fq^*$. If $b=0$, it follows from Theorem \ref{th4} that $w(c(a,b))=\frac{(q-1)(q^{m-1}+q^{h-1})}{2}$ for each $a\in \ff_{q^{h}}^*$. If $a\neq 0$ and $b\neq 0$, then
    \[\begin{split}
    w(c(a,b))&=n-\sum_{\ell=0}^{n-1}\frac{1}q\sum_{y\in \Fq}\zeta_p^{\text{Tr}_p^q( y\text{Tr}_q^{q^h}(a\beta^\ell) +yb) }\\
    &=n-\frac{1}q\sum_{y\in \Fq}\zeta_p^{\text{Tr}_p^q(yb)} \sum_{\ell=0}^{n-1}\zeta_p^{\text{Tr}_p^{q^h}(ay\beta^\ell )}\\
    &=n-\frac{1}q\sum_{y\in \Fq}\zeta_p^{\text{Tr}_p^q(yb) } \sum_{\ell_2=0}^{q^h-2}\sum_{\ell_1=0}^{\frac{q^h-1}2}\zeta_p^{\text{Tr}_p^{q^h}(ay\beta^{[(q^h-1)\ell_1+\ell_2]} )}.
    \end{split}\]
    Note that $\text{ord}(\beta)=q^h-1$, we have
    \[\begin{split}
    w(c(a,b))=n-\frac{q^h+1}{2q}\sum_{y\in \Fq}\zeta_p^{\text{Tr}_p^q(yb) } \sum_{\ell_2=0}^{q^h-2}\zeta_p^{\text{Tr}_p^{q^h}(ay\beta^{\ell_2})}\\
    =n-\frac{q^h+1}{2q}\sum_{y\in \Fq}\zeta_p^{\text{Tr}_p^q(yb) } \sum_{x\in \ff_{q^h}^*}\zeta_p^{\text{Tr}_p^{q^h}(ayx)}\\
    =n-\frac{q^h+1}{2q}\sum_{y\in \Fq}\zeta_p^{\text{Tr}_p^q(yb) } \sum_{x\in \ff_{q^h}}\zeta_p^{\text{Tr}_p^{q^h}(ayx)}
    \end{split}\]
    By using the orthogonality relations for additive characters, we have
    \[\begin{split}
    w(c(a,b))=n-\frac{q^h+1}{2q} q^h=\frac{q^m-q^{m-1}-q^{h-1}-1}2.
    \end{split}\]
 
    This completes the proof.
\end{IEEEproof}\

\begin{table}
    \caption{THE WEIGHT DISTRIBUTION OF $\mathcal{C}_{(q,m,2,\delta_1)}$ WHEN $m$ IS ODD}
    \begin{center}
        \begin{tabular}{cc}
            \toprule
            Weight & Frequency\\
            \midrule
            $0$ & $1$\\
            $ \frac{q^m-q^{m-1}-q^{\frac{m-1}2}-1}2$ & $\frac{(q-1)(q^m-1)}{2}$\\
            $\frac{q^m-q^{m-1}}2$ & $q^m-1$\\
            $\frac{q^m-q^{m-1}+q^{\frac{m-1}2}-1}2$ & $\frac{(q-1)(q^m-1)}{2}$\\
            $\frac{q^m-1}2$ & $q-1$\\
            \bottomrule
        \end{tabular}
    \end{center}
    \label{Tab2}
    \caption{THE WEIGHT DISTRIBUTION OF $\mathcal{C}_{(q,m,2,\delta_1)}$ WHEN $m$ IS EVEN}
    \begin{center}
        \begin{tabular}{cc}
            \toprule
            Weight & Frequency\\
            \midrule
            $0$ & $1$\\
            $ \frac{q^m-q^{m-1}-q^{\frac{m}2-1}-1}{2}$ & $\left( q-1\right) \left( q^{\frac{m}2}-1\right) $\\
            $\frac{(q-1)(q^{m-1}+q^{\frac{m}2-1})}{2}$ & $q^{\frac{m}2}-1$\\
            $\frac{q^m-1}2$ & $q-1$\\
            \bottomrule
        \end{tabular}
    \end{center}
    \label{Tab3}
\end{table}

\begin{example}
    When $(q,m)=(3,3)$, the BCH code $\cC_{(q,m,2,\delta_1)}$ is a $[13,4,7]$ code over $\ff_3$ with weight enumerator $1+26z^7+26z^9+26z^{10}+2z^{13}$. It has the same parameters with the best known linear code in the Datebase.
\end{example}

\begin{example}
    When $(q,m)=(5,3)$, the BCH code $\cC_{(q,m,2,\delta_1)}$ is a $[62,4,47]$ code over $\ff_5$ with weight enumerator $1+248z^{47}+124z^{50}+248z^{52}+4z^{62}$. The best known linear code over $\ff_5$ with length $62$ and dimension $4$ has minimum distance $48$.
\end{example}

   Let $m\geq 3$ be odd and $\alpha$ be a primitive element of $\Fqm$, then $\alpha^2$ is a primitive $n$-th root of unity in $\Fqm$. From Lemmas \ref{l9} and \ref{l10}, the code $\widehat{\cC}_{(q,m,2,\delta_2)}$ has two nonzeros, and $\alpha^{2\delta_1}$ and $\alpha^{2\delta_2}$ are two non-conjugate roots of its parity-check polynomial. Let $\rho_1=q^{m-1}+q^{\frac{m-1}2}$ and $\rho_2=q^{m-1}+q^{\frac{m+1}2}$. By Lemma \ref{l3},
   \[\begin{split}
    \widehat{\cC}_{(q,m,2,\delta_2)}=\left\lbrace\left( \text{Tr}_q^{q^m}( a\alpha^{\ell\rho_1}+b\alpha^{\ell\rho_2}) \right)_{\ell=0}^{n-1}: a,b\in \Fqm \right\rbrace.
    \end{split}\]
   Note that $\text{Tr}_q^{q^m}(a \alpha^{\ell \rho_1} )=\text{Tr}_q^{q^m}(a^{q^{\frac{m+1}2}} \alpha^{q^{\frac{m+1}2}\ell \rho_1} )$
   and $q^{\frac{m+1}2} \rho_1 \equiv q^{\frac{m-1}2}+1~(\text{mod }q^m-1)$, we have
   \[\begin{split}
   \text{Tr}_q^{q^m}(a \alpha^{\ell\rho_1 } )=\text{Tr}_q^{q^m}(a^{q^{\frac{m+1}2}} \alpha^{(q^{\frac{m-1}2}+1)\ell} ). \end{split}  \]
   Similarly, we have $\text{Tr}_q^{q^m}( b \alpha^{ \ell \rho_2} ) =\text{Tr}_q^{q^m}( b^{q^{\frac{m-1}2}}\alpha^{(q^{\frac{m-3}2}+1)\ell} )$. Hence, $\widehat{\cC}_{(q,m,2,\delta_2)}$ has the same weight distribution with the following code $\mathcal{V}_1=\left\lbrace v_1(a,b): a,b\in \Fqm\right\rbrace$, where
   \[\begin{split}
    v_1(a,b)=\left( \text{Tr}_q^{q^m}(a\alpha^{(q^{\frac{m-1}2}+1)\ell }+b\alpha^{(q^{\frac{m-3}2}+1) \ell}) \right)_{\ell=0}^{n-1}.
   \end{split}\]
   Clearly, $w(v_1(a,b))=0$ for $(a,b)=(0,0)$. If $(a,b)\in \Fqm^2\backslash \{
(0,0) \}$, then
    \[\begin{split}
    w(v_1(a,b))=n-\sum_{\ell=0}^{n-1} \frac{1}q \sum_{y\in \Fq}\zeta_p^{\text{Tr}^q_p(y \text{Tr}_q^{q^m}( a\alpha^{(q^{\frac{m-1}2}+1)\ell}+b\alpha^{(q^{\frac{m-3}2}+1)\ell })) }.
    \end{split}\]
  Note that both $q^{\frac{m-1}2}+1$ and $q^{\frac{m-3}2}+1$ are even integers, then we have
  \[\begin{split}
  \sum_{\ell=0}^{2n-1}\zeta_p^{\text{Tr}^q_p(y \text{Tr}_q^{q^m}( a\alpha^{(q^{\frac{m-1}2}+1) \ell}+b\alpha^{(q^{\frac{m-3}2}+1)\ell })) }\\
  =2\sum_{\ell=0}^{n-1}\zeta_p^{\text{Tr}^q_p(y \text{Tr}_q^{q^m}( a\alpha^{(q^{\frac{m-1}2}+1)\ell }+b\alpha^{(q^{\frac{m-3}2}+1) })\ell) }. 
  \end{split} \]
  It follows that
   \[\begin{split}
  w(v_1(a,b))&=n-\frac{1}{2q} \sum_{y\in \Fq}\sum_{x\in \Fqm^*}\zeta_p^{\text{Tr}_p^q(yQ_{a,b}(x)) }\\
  &=n-\frac{q^m-1}{2q}-\frac{1}{2q}\sum_{y\in \Fq^*} \sum_{x\in \Fqm^*}\zeta_p^{\text{Tr}_p^q(yQ_{a,b}(x)) } \\
  &=\frac{(q-1)n}{q}-\frac{1}{2q} \sum_{y\in \Fq^*}\left( \sum_{x\in \Fqm}\zeta_p^{\text{Tr}_p^q(yQ_{a,b}(x)) }-1\right)\\
  &=\frac{(q-1)q^{m-1}}2-\frac{1}{2q} \sum_{y\in \Fq^*}\sum_{x\in \Fqm}\zeta_p^{\text{Tr}_p^q(yQ_{a,b}(x)) }, 
  \end{split}\]
  where $Q_{a,b}(x)=\text{Tr}_q^{q^m}( ax^{q^{\frac{m-1}2}+1 }+bx^{q^{\frac{m-3}2}+1})$.

Clearly, $Q_{a,b}(x)$ is a quadratic form in $m$ variables over $\mathbb{F}_q$, and
\[\begin{split}
Q_{a,b}(x+y)-Q_{a,b}(x)-Q_{a,b}(y)=\text{Tr}_q^{q^m}( g_{a,b}(x) \cdot y ),
\end{split}\]
where 
\[\begin{split}
g_{a,b}(x)=b^{q^{\frac{m+3}2}}x^{q^{\frac{m+3}2}}+a^{q^{\frac{m+1}2}}x^{q^{\frac{m+1}2}}+ax^{q^{\frac{m-1}2}}+bx^{q^{\frac{m-3}2}}.
\end{split}\]

Assume the rank of $Q_{a,b}(x)$ is $r_{a,b}$, then $r_{a,b}=r$ if and only if $g_{a,b}(x)=0$ has $q^{m-r}$ solutions in $\Fqm$. The number of solutions of the above equation equals the number of solutions of the following equation
\[\begin{split}
b^{q^{\frac{m+3}2}}x^{q^{3}}+a^{q^{\frac{m+1}2}}x^{q^{2}}+ax^{q}+bx=0,
\end{split}\]
which has at most $q^3$ solutions. Thus, we have the following result.

\begin{lemma}\label{l12}
    For $(a,b)\in \Fqm^2\backslash \{ (0,0)\}$, let $r_{a,b}$ be the rank of $Q_{a,b}(x)$.
    \begin{itemize}
        \item [{\rm (\romannumeral1)}] If $m=3$, the possible values of $r_{a,b}$ are $m$, $m-1$ and $m-2$.
        \item [{\rm (\romannumeral2)}] If $m\geq 5$ is odd, the possible values of $r_{a,b}$ are $m$, $m-1$, $m-2$ and $
        m-3$.
    \end{itemize}
\end{lemma}

By Lemma \ref{l4}, we have
\begin{equation}\label{EQ7}
\begin{split}
w(v_1(a,b))=\frac{(q-1)q^{m-1}}{2}-\frac{T(a,b)}{2q} \sum_{y\in\Fq^*}\eta(y^{r_{a,b}}),
\end{split}
\end{equation}
where $T(a,b)=\sum_{x\in \Fqm}\zeta_p^{\text{Tr}^q_p(Q_{a,b}(x)) }$. In order to determine the value distribution of
$T(a,b)$, we need the following results on moments of $T(a,b)$.

\begin{lemma}\label{l13}
    Let $m$ be odd and $T(a,b)$ be defined as above, and let $S(a,b)=\sum_{y\in \Fq^*}T(ay,by)$.
    \begin{itemize}
        \item [{\rm (\romannumeral1)}] $\sum_{a,b\in \Fqm}T(a,b)=q^{2m}$.
        \item [{\rm (\romannumeral2)}] If $q\equiv 3~({\rm mod}~4)$, then $\sum_{a,b\in \Fqm}T(a,b)^2=q^{2m}$. If $q\equiv 1~({\rm mod}~4)$, then
        \[\begin{split}
        \sum_{a,b\in \Fqm}T(a,b)^2=(2q^m-1)q^{2m}.
        \end{split}\]
        \item [{\rm (\romannumeral3)}] $\sum_{a,b\in \Fqm}T(a,b)^3=[q^m+q^{m-1}-1]q^{2m+1}$.
        \item [{\rm (\romannumeral4)}] $\sum_{a,b\in \Fqm}S(a,b)^2=(q-1)^2q^{3m}$.
    \end{itemize}
\end{lemma}

\begin{IEEEproof}
    (\romannumeral1) The identity is trivially true. 
   
    (\romannumeral2) We observe that
    \[\begin{split}
    &\sum_{a,b\in \mathbb{F}_{q^m}}T(a,b)^2=\sum_{a,b\in \mathbb{F}_{q^m}}\sum_{x,y\in \Fqm}\zeta_p^{\text{Tr}^q_p(Q_{a,b}(x)+Q_{a,b}(y))}\\
    &=\sum_{x,y\in \Fqm}\sum_{a\in \Fqm}\zeta_p^{\text{Tr}_p^{q^m}(af_m(x,y))}\sum_{b\in \Fqm}\zeta_p^{\text{Tr}_p^{q^m}(bf_{m-2}(x,y))}\\
    &=A\cdot q^{2m},
    \end{split}\]
    where $f_i(x,y)=x^{q^{\frac{i-1}2}+1}+y^{q^{\frac{i-1}2}+1}$ for every positive odd $i$ and $A$ denotes the number of the pair $(x,y)\in \Fqm^2$, which is a solution of the following system of equations:
    \[\begin{split}
    \begin{cases}
    x^{q^{\frac{m-1}2}+1}+y^{q^{\frac{m-1}2}+1}=0,\\
    x^{q^{\frac{m-3}2}+1}+y^{q^{\frac{m-3}2}+1}=0.
    \end{cases}
    \end{split}\]
    
    Clearly, $(0,0)$ is a solution of the above system of equations. If $y\neq 0$, the system above is equivalent to
    \begin{equation}\label{EQ8}
    \begin{split}
    \begin{cases}
    (\frac{x}y)^{q^{\frac{m-1}2}+1} =-1,\\
    (\frac{x}y)^{q^{\frac{m-3}2}+1} =-1.\\
    \end{cases}
    \end{split}
    \end{equation}
    Let $B$ be the number of pair $(x,y)\in \Fqm^2$, which is a solution of the system of equations (\ref{EQ8}). Clearly, $A=B+1$. Thus, it suffices to determine the value of $B$. 
    
    Now assume the system of equations (\ref{EQ8}) has a solution $(x,y)$. Note that $\gcd(2(q^{\frac{m-1}2}+1),2(q^{\frac{m-3}2}+1))=4$, from (\ref{EQ8}), ${\rm ord}(\frac{x}y)$ divides $4$. We claim ${\rm ord}(\frac{x}y)=4$. Otherwise, we have $1=(\frac{x}y)^{q^{\frac{m-1}2}+1}=-1$, which is impossible. Thus, if the system of equations (\ref{EQ8}) has solutions, then $q^{\frac{m-1}2}+1\equiv q^{\frac{m-3}2}+1\equiv2~({\rm mod}~4)$, which deduces $q\equiv 1~({\rm mod}~4)$. Now assume $q\equiv 1~({\rm mod}~4)$, then the system (\ref{EQ8}) is equivalent to $(\frac{x}y)^2=-1 $. Thus, $B=2(q^m-1)$. Note that $B=0$ if $q\equiv 3~({\rm mod}~4)$. The result follows. 

    (\romannumeral3) Similar to (\romannumeral2), we have
    \[\begin{split}
    \sum_{a,b\in \mathbb{F}_{q^m}}T(a,b)^3=M\cdot q^{2m},
    \end{split}\]
   where $M$ is the number of triple $(x,y,z)\in \Fqm^3$, which is a solution of the following system of equations:
   \[\begin{split}
   \begin{cases}
   x^{q^{\frac{m-1}2}+1}+y^{q^{\frac{m-1}2}+1}+z^{q^{\frac{m-1}2}+1}=0,\\
   x^{q^{\frac{m-3}2}+1}+y^{q^{\frac{m-3}2}+1}+z^{q^{\frac{m-3}2}+1}=0.
   \end{cases}
   \end{split}\]
  From (\romannumeral2), the number of triples $(x,y,0)\in \Fqm^3$ which are solutions of the above system of equations is equal to $A$. If $z\neq 0$, the above system is equivalent to 
   \begin{numcases}{}
   (\frac{x}z)^{q^{\frac{m-1}2}+1}+ (\frac{y}z)^{q^{\frac{m-1}2}+1}+1=0,\label{EQ9}\\
   (\frac{x}z)^{q^{\frac{m-3}2}+1}+(\frac{y}z)^{q^{\frac{m-3}2}+1}+1=0.\label{EQ10}
    \end{numcases}
    
    Assume the above system of equations has a solution $(x,y,z)\in \Fqm^3$. Raising to the $q$-th power both sides of (\ref{EQ10}), we have  
   \begin{equation}\label{EQ11}
   \begin{split}
    (\frac{x}z)^{q^{\frac{m-1}2}+q}+ (\frac{y}z)^{q^{\frac{m-1}2}+q}+ 1=0.
   \end{split}
   \end{equation}
    Taking the difference between (\ref{EQ11}) and (\ref{EQ9}), we obtain
    \begin{equation}
    \begin{split}
   (\frac{x}z)^{q^{\frac{m-1}2}+1}[(\frac{x}z)^{q-1}-1]+ (\frac{y}z)^{q^{\frac{m-1}2}+1}[(\frac{y}z)^{q-1}-1]=0.\label{EQ12}
    \end{split}
    \end{equation}
    From (\ref{EQ12}), if one of $\frac{x}z, \frac{y}z$ is in $\ff_q$, then the other must also be in $\mathbb{F}_q$. We claim both $\frac{x}z$ and $\frac{y}z$ are in $\ff_q$. Otherwise, we will gives a contradiction. On the one hand, it follows from (\ref{EQ12}) that 
    \[\begin{split}
    (\frac{x}y)^{q^{\frac{m-1}2}+1}=\frac{z^{q-1}-y^{q-1}}{x^{q-1}-z^{q-1}}.
    \end{split} \] 
    On the other hand, notice that both $\frac{x}z$ and $\frac{y}z$ are in $\Fqm$. Raising to the $q^{\frac{m+3}2}$-th power both sides of (\ref{EQ9}) and (\ref{EQ10}), we have
    \[\begin{split}
     \begin{cases}
    (\frac{x}z)^{q^{\frac{m+3}2}+1}+(\frac{y}z)^{q^{\frac{m+3}2}+1}+1=0,\\
     (\frac{x}z)^{q^{\frac{m+3}2}+q}+(\frac{y}z)^{q^{\frac{m+3}2}+q}+1=0,
    \end{cases}
    \end{split}\]
    which deduces
    \[\begin{split}
    (\frac{x}z)^{q^{\frac{m+3}2}+1}[(\frac{x}z)^{q-1}-1]+(\frac{y}z)^{q^{\frac{m+3}2}+1}[(\frac{y}z)^{q-1}-1]=0.
    \end{split}\]
    This gives that
     \[\begin{split}
    (\frac{x}y)^{q^{\frac{m+3}2}+1}=\frac{z^{q-1}-y^{q-1}}{x^{q-1}-z^{q-1}}.
    \end{split} \] 
    Therefore, $(\frac{x}y)^{q^{\frac{m-1}2}+1}=(\frac{x}y)^{q^{\frac{m+3}2}+1} $. It follows that
    \[\begin{split}
    (\frac{x}y)^{q^{\frac{m+3}2}-q^{\frac{m-1}2}}=1,
    \end{split}\]
    since $\frac{x}y \notin \ff_q$. Notice that $\frac{x}y\in \Fqm$ and $\gcd(q^{\frac{m+3}2}-q^{\frac{m-1}2}, q^m-1)=q-1$, we obtain $(\frac{x}y)^{q-1}=1$. Now we assume $x=\sigma y$, where $\sigma\in \ff_q^*$. Taking it into the equations (\ref{EQ9}) and (\ref{EQ10}), we have
    \[\begin{split}
    \begin{cases}
    (\sigma^2+1)(\frac{y}z)^{q^{\frac{m-1}2}+1}+1=0,\\
    (\sigma^2+1)(\frac{y}z)^{q^{\frac{m-3}2}+1}+1=0.
    \end{cases}
    \end{split}\]
   
    Obviously, both $\sigma^2+1$ and $\frac{y}z$ are nonzero elements. Hence, $(\frac{y}z)^{q^{\frac{m-1}2}+1}=(\frac{y}z)^{q^{\frac{m-3}2}+1}$. Note that $\frac{y}z\in \Fqm$, it follows that $(\frac{y}z)^{q-1}=1$, which contradicts the fact that $\frac{y}z\notin \ff_q$. Therefore, the equations (\ref{EQ9}) and (\ref{EQ10}) have a solution $(x,y,z)\in \Fqm^3$, then both $\frac{x}z$ and $\frac{y}z$ are in $\ff_q$. Let $ D=\left|\left\lbrace(x,y)\in \mathbb{F}_{q}^2: x^2+y^2+1=0  \right\rbrace \right|$, then the number of triples $(x,y,z)\in \Fqm^3$ which are solutions of (\ref{EQ9}) and (\ref{EQ10}) is equal to $(q^m-1)D$.
     
     Noticing the quadratic equation over $\ff_q$ has been studied by Wan, as an application of his results (see \cite[Ch. 1, Th. 1.27 ]{wan2002geometry}), we have $D=q-1$ if $q\equiv 1~ ({\rm mod}~4)$, and $D=q+1$ if $q\equiv 3 ~({\rm mod}~4)$. Note that $M=A+(q^m-1)D$, the result follows. 
 
    The proof of (\romannumeral4) is very similar to that of (\romannumeral2), and thus is omitted here.
\end{IEEEproof}\

Let $m\geq 3$ be odd. According to Lemma \ref{l4}, if $q\equiv
1~(\text{mod}~4)$, for $\epsilon=\pm 1$ and $0\leq
i\leq 3$, we define that 
\[\begin{split}
N_{\epsilon, i}=\left\lbrace (a,b)\in
\mathbb{F}_{q^m}^2\backslash \left\lbrace (0,0)\right\rbrace  :
T(a,b)=\epsilon q^{\frac{m+i}2}\right\rbrace.
\end{split}\]

If $q\equiv 3~({\rm
mod}~4)$, for $\epsilon=\pm 1$ and $i\in \{ 0,2\}$, we define that
\[\begin{split}
N_{\epsilon, i}=\left\lbrace (a,b)\in \mathbb{F}_{q^m}^2\backslash
\left\lbrace (0,0)\right\rbrace  : T(a,b)=\epsilon q^{\frac{m+i}2}\sqrt{-1}\right\rbrace .
\end{split}\]
For $i\in \{ 1,3\}$,
define
\[\begin{split}
N_{\epsilon, i}=\left\lbrace (a,b)\in \mathbb{F}_{q^m}^2\backslash \left\lbrace (0,0)\right\rbrace  : T(a,b)=\epsilon q^{\frac{m+i}2}\right\rbrace.
\end{split}\]
And $n_{\epsilon, i}=\left| N_{\epsilon, i} \right|$.

\begin{lemma}\label{l14}
    Let $m\geq 3$ be odd, then the value distribution of $T(a,b)$ is listed in Table \ref{Tab4}.
\end{lemma}

\begin{IEEEproof}
    We choose an element $\omega\in \mathbb{F}_{q}^*$ such that $\eta(\omega)=-1$, where $\eta$ is the quadratic character of $\ff_q$. When $i\in \{ 0,2\}$, for any $(a,b)\in N_{1,i}$, from Lemma \ref{l4}, we have $T(\omega a, \omega b)=-T(a,b)$, since $m-i$ is odd. Then the map $(a,b)\mapsto (\omega a,\omega b)$ gives a $1$-to-$1$ correspondence from $N_{1,i}$ to $N_{-1,i}$. Thus, $n_{1,0}=n_{-1,0}$ and $n_{1,2}=n_{-1,2}$. It follows from Lemma \ref{l13} that
     \[\begin{split}
     &\sum_{a,b\in \Fqm}T(a,b)\\
     &=q^m+(n_{1,1}-n_{-1,1})q^{\frac{m+1}2}+(n_{1,3}-n_{-1,3})q^{\frac{m+3}2}\\
     &=q^{2m},
    \end{split}\]
    \[\begin{split}
     &\sum_{a,b\in \Fqm}T(a,b)^3\\
     &=q^{3m}+(n_{1,1}-n_{-1,1})q^{\frac{3m+3}2}+(n_{1,3}-n_{-1,3})q^{\frac{3m+9}2}\\
     &=[q^m+q^{m-1}-1]q^{2m+1}.
    \end{split}\]
    It deduces that $n_{1,1}-n_{-1,1}=(q^m-1)q^{\frac{m-1}2}$ and $n_{1,3}=n_{-1,3}$. If $n_{1,3}>0$, from (\ref{EQ7}), we have
    \[\begin{split}
    w(v_1(a,b))=\frac{(q-1)q^{m-1}}{2}-\frac{(q-1)q^{\frac{m+1}2}}{2}<\delta_2+1.
    \end{split}\]
    However, from Lemma \ref{l1}, the minimum distance of $\widehat{\mathcal{C}}_{(q,m,2,\delta_2)}$ is at least $\delta_2+1$. Hence, $n_{1,3}=0$. That is, $n_{1,3}=n_{-1,3}=0$. At this point, from (\romannumeral4) of Lemma \ref{l13}, 
     \[\begin{split}
    \sum_{a,b\in \Fqm}S(a,b)^2&=(q^2-1)[q^{2m}+(n_{1,1}+n_{-1,1})q^{m+1}]\\
    &=(q^2-1)q^{3m}.
    \end{split}\]
    
   Notice that if $q\equiv 1~({\rm mod}~4)$, 
   \[\begin{split}
   &\sum_{a,b\in \mathbb{F}_{q^m}}T(a,b)^2\\
   =&q^{2m}+2n_{1,0}q^m+(n_{1,1}+n_{-1,1})q^{m+1}+2n_{1,2}q^{m+2}\\
   =&(2q^m-1)q^{2m}.
   \end{split}\]
   If $q\equiv 3~({\rm mod}~4)$, then
     \[\begin{split}
     &\sum_{a,b\in \mathbb{F}_{q^m}}T(a,b)^2\\
     =&q^{2m}-2n_{1,0}q^m+(n_{1,1}+n_{-1,1})q^{m+1}-2n_{1,2}q^{m+2}\\
     =&q^{2m}.
   \end{split}\]
    Moreover, $1+2n_{1,0}+n_{1,1}+n_{-1,1}+2n_{1,2}=q^{2m}$. Simplifying the above equations leads to
    \[\begin{split}
      \begin{cases}n_{1,1}-n_{-1,1}=(q^m-1)q^{\frac{m-1}2},\\
    n_{1,1}+n_{-1,1}=(q^m-1)q^{m-1},\\
    2n_{1,0}+2q^2n_{1,2}=(q^m-1)q^{m},\\
    2n_{1,0}+2n_{1,2}=(q^m-1)(q^{m}-q^{m-1}+1).
    \end{cases}
    \end{split}\]
   
    The value distribution of $T(a,b)$ then follows.
\end{IEEEproof}\

\begin{table}
    \caption{ THE VALUE DISTRIBUTION OF $T(a,b)$ }
    \begin{center}
        \begin{tabular}{ccc}
            \toprule
            Rank~$r_{a,b}$& Value $T(a,b)$& Multiplicity\\
            \midrule
            $m$ & $q^{\frac{m}2} \sqrt{(-1)^{\frac{q-1}2}} $& $\frac{(q^m-1)(q^{m+2}-q^{m+1}-q^m+q^2)}{2(q^2-1)}$\\
            $m$ & $-q^{\frac{m}2} \sqrt{(-1)^{\frac{q-1}2}}$&$\frac{(q^m-1)(q^{m+2}-q^{m+1}-q^m+q^2)}{2(q^2-1)}$\\
            $m-1$ & $q^{\frac{m+1}2}$&$\frac{(q^m-1)(q^{m-1}+q^{\frac{m-1}2})}{2}$\\
            $m-1$ & $-q^{\frac{m+1}2}$&$\frac{(q^m-1)(q^{m-1}-q^{\frac{m-1}2})}{2}$\\
            $m-2$ & $q^{\frac{m+2}2}\sqrt{(-1)^{\frac{q-1}2}}$&$\frac{(q^m-1)(q^{m-1}-1)}{2(q^2-1)}$\\
            $m-2$ & $-q^{\frac{m+2}2}\sqrt{(-1)^{\frac{q-1}2}}$&$\frac{(q^m-1)(q^{m-1}-1)}{2(q^2-1)}$\\
            $0$ & $q^m$&$1$\\
            \bottomrule
        \end{tabular}
    \end{center}
    \label{Tab4}
\end{table}

Let $m\geq 2$ be even and $\alpha$ be a primitive element of $\mathbb{F}_{q^m}$, then $\alpha^2$ is a primitive $n$-th root of unity in $\mathbb{F}_{q^m}$. From Lemmas \ref{l9} and \ref{l10}, the code $\widehat{\mathcal{C}}_{(q,m,2,\delta_2)}$ has two nonzeros, and $\alpha^{2\delta_1}$ and $\alpha^{2\delta_2}$ are two non-conjugate roots of its parity-check polynomial. Let $h=\frac{m}2$, $\rho_1=q^{m-1}+q^{h-1}$ and $\rho_2=q^{m-1}+q^h$. By Lemma \ref{l3},
\[\begin{split}
\widehat{\mathcal{C}}_{(q,m,2,\delta_2)}=\left\lbrace c(a,b): a\in\mathbb{F}_{q^h}, b\in \mathbb{F}_{q^m}\right\rbrace,
\end{split}\]
where  
\[\begin{split}
c(a,b)=\left( \text{Tr}_q^{q^h}( a\alpha^{\ell\rho_1})+\text{Tr}_q^{q^{m}}(b\alpha^{\ell\rho_2 }) \right)_{\ell=0}^{n-1}.
\end{split}\]

Since $\text{Tr}_q^{q^h}( a\alpha^{\ell\rho_1})=\text{Tr}_q^{q^h}( a^q\alpha^{(q^h+1)\ell})$ and
\[\begin{split}
\text{Tr}_q^{q^m}(b\alpha^{\ell\rho_2 })=\text{Tr}_q^{q^m}(b^{q^h}\alpha^{(q^{h-1}+1)\ell}),
\end{split}\]
it follows that $\widehat{\mathcal{C}}_{(q,m,2,\delta_2)}$ has the same weight distribution with the code
\[\begin{split}
\mathcal{V}_2=\left\lbrace v_2(a,b): a\in\mathbb{F}_{q^h}, b\in \mathbb{F}_{q^m} \right\rbrace,
\end{split}\]
where 
\[\begin{split}
v_2(a,b)=\left( \text{Tr}_q^{q^h}( a\alpha^{(q^h+1)\ell}) +\text{Tr}_q^{q^m}( b\alpha^{(q^{h-1}+1)\ell })\right)_{\ell=0}^{n-1}.
\end{split} \]

Clearly, $w(v_2(a,b))=0$ for $(a,b)=(0,0)$. If $(a,b)\in\mathbb{F}_{q^h}\times\mathbb{F}_{q^m}\backslash \{ (0,0)
\}$, similar to the odd case $m\geq 3$, we have 
\[\begin{split}
w(v_2(a,b))&=\frac{(q-1)q^{m-1}}{2}-\frac{1}{2q}\sum_{y\in \ff_q^*}\sum_{x\in \Fqm}\zeta_p^{\text{Tr}^q_p(y\overline{Q}_{a,b}(x))}\\
&=\frac{(q-1)q^{m-1}}{2}-\frac{\overline{T}(a,b)}{2q}\sum_{y\in \ff_q^*}\eta(y^{r_{a,b}}),
\end{split}\]
where $\overline{Q}_{a,b}(x)= \text{Tr}_q^{q^h}(
ax^{q^h+1})+\text{Tr}_q^{q^m}(
bx^{q^{h-1}+1})$, $r_{a,b}$ is the rank of $\overline{Q}_{a,b}(x)$ and
\[\begin{split}
\overline{T}(a,b)=\sum_{x\in \Fqm}\zeta_p^{\text{Tr}^q_p(\overline{Q}_{a,b}(x))}.
\end{split}\]
Clearly, in order to determine the weight of $v_2(a,b)$, it suffices to determine the value distribution of $\overline{T}(a,b)$. Fortunately, the value distribution of $\overline{T}(a,b)$ was determine in \cite{luo2009exponential}, and we list it in Table \ref{Tab5} .

\begin{table}
	\caption{ THE VALUE DISTRIBUTION OF $\overline{T}(a,b)$ }
	\begin{center}
		\begin{tabular}{ccc}
			\toprule
			Rank~$r_{a,b}$& Value $\overline{T}(a,b)$& Multiplicity\\
			\midrule
			$m$ & $q^{\frac{m}2}$& $\frac{(q^{m}-1)(q^{\frac{m+2}2}+q)}{2(q+1)}$\\
			$m$ & $-q^{\frac{m}2}$&$\frac{(q^{\frac{m}2}-1)(q^{m+1}-2q^{m}+q)}{2(q-1)}$\\
			$m-1$ & $q^{\frac{m+1}2}\sqrt{(-1)^{\frac{q-1}2}}$&$\frac{(q^m-1)q^{\frac{m-2}2}}{2}$\\
			$m-1$ & $-q^{\frac{m+1}2}\sqrt{(-1)^{\frac{q-1}2}}$&$\frac{(q^m-1)q^{\frac{m-2}2}}{2}$\\
			$m-2$ & $-q^{\frac{m+2}2}$&$\frac{(q^m-1)(q^{\frac{m-2}2}-1)}{q^2-1}$\\
			$0$ & $q^m$&$1$\\
			\bottomrule
		\end{tabular}
	\end{center}
	\label{Tab5}
\end{table}

\begin{theorem}\label{th6}
    The BCH code $\widehat{\mathcal{C}}_{(q,m,2,\delta_2)}$ has parameters $[\frac{q^m-1}2,k,d]$, where
    \begin{itemize}
        \item [{\rm (\romannumeral1)}] if $m$ is odd, then $k=2m$, $d=\frac{(q-1)(q^{m-1}-q^{\frac{m-1}2})}{2}$ and $\widehat{\mathcal{C}}_{(q,m,2,\delta_2)}$ is a there-weight code. In addition, the weight distribution of $\widehat{\mathcal{C}}_{(q,m,2,\delta_2)}$ is listed in Table \ref{Tab7}.
        \item [{\rm (\romannumeral2)}] if $m$ is even, then $k=\frac{3m}2$, $d=\frac{(q-1)(q^{m-1}-q^{\frac{m-2}2})}{2}$ and $\widehat{\mathcal{C}}_{(q,m,2,\delta_2)}$ is a four-weight code for $m\geq 4$. In addition, the weight distribution of $\widehat{\mathcal{C}}_{(q,m,2,\delta_2)}$ is listed in Table \ref{Tab8}.
    \end{itemize}
\end{theorem}

\begin{IEEEproof}
	We only prove the case that $m\geq3$ is odd, and the even case is similar. If $r_{a,b}$ is odd, from (\ref{EQ7}), we have 
	\[\begin{split}
	w(v_1(a,b))=\frac{(q-1)(q^m-1)}2.
	\end{split}\]
    Hence, the number of such codewords is equal to $n_{1,0}+n_{-1,0}+n_{1,2}+n_{-1,2}$. If $r_{a,b}$ is even, from (\ref{EQ7}), then $w(v_1(a,b))=\frac{(q-1)q^{m-1}}{2}-\frac{(q-1)T(a,b)}{2q}$.
    By Lemma \ref{l14}, the weight distribution of the code then follows.
\end{IEEEproof}

\begin{table}
    \caption{THE WEIGHT DISTRIBUTION OF $\widehat{\mathcal{C}}_{(q,m,2,\delta_2)}$ WHEN $m$ IS ODD}
    \begin{center}
        \begin{tabular}{cc}
            \toprule
            Weight & Frequency\\
            \midrule
            $0$ & $1$\\
            $\frac{(q-1)(q^{m-1}-q^{\frac{m-1}2})}{2}$ & $\frac{(q^m-1)(q^{m-1}+q^{\frac{m-1}2})}{2}$\\
            $\frac{(q-1)q^{m-1}}{2}$ & $\left( q^m-1\right) \left( q^m-q^{m-1}+1\right) $\\
            $\frac{(q-1)(q^{m-1}+q^{\frac{m-1}2})}2$ & $\frac{(q^m-1)(q^{m-1}-q^{\frac{m-1}2})}{2}$\\
            \bottomrule
        \end{tabular}
    \end{center}
    \label{Tab7}
 \end{table}
   \begin{table}   
    \caption{THE WEIGHT DISTRIBUTION OF $\widehat{\mathcal{C}}_{(q,m,2,\delta_2)}$ WHEN $m$ IS EVEN}
    \begin{center}
        \begin{tabular}{cc}
            \toprule
            Weight & Frequency\\
            \midrule
            $0$ & $1$\\
            $\frac{(q-1)(q^{m-1}-q^{\frac{m-2}2})}2$ & $\frac{(q^{m}-1)(q^{\frac{m}2+1}+q)}{2(q+1)}$\\
            $\frac{(q-1)q^{m-1}}{2}$ & $q^{\frac{m-2}2}(q^m-1)$\\
            $\frac{(q-1)(q^{m-1}+q^{\frac{m-2}2})}{2}$ & $\frac{(q^{\frac{m}2+1}-q)(q^m-2q^{m-1}+1)}{2(q-1)} $\\
            $\frac{(q-1)(q^{m-1}+q^{\frac{m}2})}2$ & $\frac{(q^m-1)(q^{\frac{m-2}2}-1)}{q^2-1}$\\
            \bottomrule
        \end{tabular}
    \end{center}
    \label{Tab8}
\end{table}

It is observed that the weight of the BCH code in Theorem \ref{th6} has a common divisor $\frac{q-1}2$. Hence, we consider a punctured code of this class of BCH codes. Let $N=\frac{q^m-1}{q-1}$. If $m\geq 3$ is odd, we define $\mathcal{V}_3=\left\lbrace v_3(a,b): a,b\in \mathbb{F}_{q^m} \right\rbrace $, where
\[\begin{split}
v_3(a,b)=\left( \text{Tr}_q^{q^m}(a\alpha^{\ell(q^{\frac{m-1}2}+1) }+b\alpha^{\ell(q^{\frac{m-3}2}+1 ) }) \right)_{\ell=0}^{N-1}.
\end{split}\]
If $m\geq 2$ is even, define 
\[\begin{split}
\mathcal{V}_3=\left\lbrace v_3(a,b): a\in\mathbb{F}_{q^{\frac{m}2}}, b\in \mathbb{F}_{q^m} \right\rbrace,
\end{split}\]
where
\[\begin{split}
v_3(a,b)=\left( \text{Tr}_q^{q^{\frac{m}2}}( a\alpha^{\ell(q^{\frac{m}2}+1) }) +\text{Tr}_q^{q^{m}}( b\alpha^{\ell(q^{\frac{m-2}2}+1) }) \right)_{\ell=0}^{N-1}.
\end{split}\]

Let $\gamma=\alpha^N$, then $\gamma$ is a primitive element of $\ff_q$. Sequentially, $\gamma^{q^i+1}=\gamma^2$ for every positive integer $i$. Let  $t=\frac{q-1}2$, we have
\[\begin{split}
v_2(a,b)=v_3(a,b)\parallel\gamma^2 v_3(a,b)\parallel\cdots \parallel\gamma^{2(t-1)}v_3(a,b),
\end{split}\]
where $v_2(a,b)$ is defined as above and $\parallel$ denotes the concatenation of vectors. Hence we obtain a punctured linear code $\mathcal{V}_3$ of the code $\mathcal{V}_2$. By Theorem \ref{th6}, we directly obtain the following result.

\begin{theorem}
    Let $\mathcal{V}_3$ be defined as above. Then $\mathcal{V}_{3}$ has parameters $[\frac{q^m-1}{q-1},k,d]$, where
    \begin{itemize}
        \item [{\rm (\romannumeral1)}] if $m$ is odd, then $k=2m$, $d=q^{m-1}-q^{\frac{m-1}2}$ and $\mathcal{V}_3$ is a there-weight code. In addition, the weight distribution of $\mathcal{V}_3$ is listed in Table \ref{Tab9}.
        \item [{\rm (\romannumeral2)}] if $m$ is even, then $k=\frac{3m}2$, $d=q^{m-1}-q^{\frac{m-2}2}$ and $\mathcal{V}_{3}$ is a four-weight code for $m\geq 4$. In addition, the weight distribution of $\mathcal{V}_{3}$ is listed in Table \ref{Tab10}.
    \end{itemize}
\end{theorem}

\begin{table}
    \caption{PUNCTURING CODE FROM $\widehat{\mathcal{C}}_{(q,m,2,\delta_2)}$ WHEN $m$ IS ODD }
    \begin{center}
        \begin{tabular}{cc}
            \toprule
            Weight & Frequency\\
            \midrule
            $0$ & $1$\\
            $q^{m-1}-q^{\frac{m-1}2}$ & $\frac{(q^m-1)(q^{m-1}+q^{\frac{m-1}2})}{2}$\\
            $q^{m-1}$ & $\left( q^m-1\right) \left( q^m-q^{m-1}+1\right) $\\
            $q^{m-1}+q^{\frac{m-1}2}$ & $\frac{(q^m-1)(q^{m-1}-q^{\frac{m-1}2})}{2}$\\
            \bottomrule
        \end{tabular}
    \end{center}
    \label{Tab9}
    \caption{PUNCTURING CODE FROM $\widehat{\mathcal{C}}_{(q,m,2,\delta_2)}$ WHEN $m$ IS EVEN }
    \begin{center}
        \begin{tabular}{cc}
            \toprule
            Weight & Frequency\\
            \midrule
            $0$ & $1$\\
            $q^{m-1}-q^{\frac{m-2}2}$ & $\frac{(q^{m}-1)(q^{\frac{m}2+1}+q)}{2(q+1)}$\\
            $q^{m-1}$ & $q^{\frac{m-2}2}(q^m-1)$\\
            $q^{m-1}+q^{\frac{m-2}2}$ & $\frac{(q^{\frac{m}2+1}-q)(q^m-2q^{m-1}+1)}{2(q-1)} $\\
            $q^{m-1}+q^{\frac{m}2}$ & $\frac{(q^m-1)(q^{\frac{m-2}2}-1)}{q^2-1}$\\
            \bottomrule
        \end{tabular}
    \end{center}
    \label{Tab10}
\end{table}

\begin{example}
    When $(q,m)=(3,3)$, the BCH code $\widehat{\mathcal{C}}_{(q,m,2,\delta_2)}$ is a $[13,6,6]$ code over $\mathbb{F}_3$ with weight enumerator $1+156z^6+494z^9+78z^{12}$. This code has the same parameters with the best known in the Datebase.
\end{example}

\begin{example}
    When $q$ is odd and $m=2$, the linear code $\mathcal{V}_3$ is a $[q+1,3,q-1]$ MDS code over $\mathbb{F}_q$ with weight enumerator $1+\frac{q(q^2-1)}2z^{w_1}+(q^2-1)z^{w_2}+\frac{q(q^2-2q+1)}2z^{w_3}$, where $w_1=q-1$, $w_2=q$, $w_3=q+1$.
\end{example}

\begin{example}
    When $(q,m)=(5,3)$, the linear code $\mathcal{V}_3$ is a $[31,6,20]$ code over $\mathbb{F}_5$ with weight enumerator $1+1860z^{20}+12524z^{25}+1240z^{30}$. This code has the same parameters with the best known in the Datebase.
\end{example}

\begin{table}
    \caption{THE WEIGHT DISTRIBUTION OF $\mathcal{C}_{(q,m,2,\delta_2)}$ WHEN $m$ IS ODD }
    \begin{center}
        \begin{tabular}{cc}
            \toprule
            Weight & Multiplicity\\
            \midrule
            $0$&$1$\\
            $\frac{q^m-q^{m-1}-q^{\frac{m+1}2}-1}{2}$& $\frac{(q^m-1)(q^{m-1}-1)}{2(q+1)}$\\
            $\frac{q^m-q^{m-1}-q^{\frac{m+1}2}+q^{\frac{m-1}2}}2$&$\frac{(q^m-1)(q^{m-1}+q^{\frac{m-1}2})}{2}$\\
            $\frac{q^m-q^{m-1}-q^{\frac{m-1}2}-1}2$&$\frac{(q^m-1)(q^{m+2}-q^{m}-q^{m-1}-q^{\frac{m+3}2}+q^{\frac{m-1}2}+q^2)}{2(q+1)}$\\
            $\frac{q^m-q^{m-1}}2$&$(q^m-1)(q^m-q^{m-1}+1)$\\
            $\frac{q^m-q^{m-1}+q^{\frac{m-1}2}-1}2$&$\frac{(q^m-1)(q^{m+2}-q^{m}-q^{m-1}+q^{\frac{m+3}2}-q^{\frac{m-1}2}+q^2)}{2(q+1)}$\\
            $\frac{q^m-q^{m-1}+q^{\frac{m+1}2}-q^{\frac{m-1}2}}2$&$\frac{(q^m-1)(q^{m-1}-q^{\frac{m-1}2})}{2}$\\
            $\frac{q^m-q^{m-1}+q^{\frac{m+1}2}-1}2$&$\frac{(q^m-1)(q^{m-1}-1)}{2(q+1)}$\\
            $\frac{q^m-1}2$&$q-1$\\
            \bottomrule
        \end{tabular}
    \end{center}
    \label{Tab11}
    \caption{THE WIGHT DISTRIBUTION OF $\mathcal{C}_{(q,m,2,\delta_2)}$ WHEN $m$ IS EVEN }
    \begin{center}
        \begin{tabular}{cc}
            \toprule
            Weight & Multiplicity\\
            \midrule
            $0$&$1$\\
            $\frac{q^m-q^{m-1}-q^{\frac{m}2}-1}{2}$& $\frac{(q^m-1)(q^{\frac{m+2}2}+q^{\frac{m-2}2}-2)}{2(q+1)}$\\
            $\frac{q^m-q^{m-1}-q^{\frac{m}2}+q^{\frac{m-2}2}}2$&$\frac{(q^m-1)(q^{\frac{m+2}2}+q)}{2(q+1)}$\\
            $\frac{q^m-q^{m-1}-q^{\frac{m-2}2}-1}2$&$\frac{(q^{\frac{m}2}-1)(q^{m+1}-2q^{m}+q)}{2}$\\
            $\frac{q^m-q^{m-1}}2$&$(q^m-1)q^{\frac{m-2}2}$\\
            $\frac{q^m-q^{m-1}+q^{\frac{m-2}2}-1}2$&$\frac{(q^m-1)(q^{\frac{m+2}2}+q)(q-1)}{2(q+1)}$\\
            $\frac{q^m-q^{m-1}+q^{\frac{m}2}-q^{\frac{m-2}2}}2$&$\frac{(q^{\frac{m+2}2}-q)(q^{m}-2q^{m-1}+1)}{2(q-1)}$\\
            $\frac{q^m-q^{m-1}+q^{\frac{m}2}-1}2$&$\frac{(q^m-1)(q^{\frac{m}2}-q^{\frac{m-2}2})}2$\\
            $\frac{q^m-q^{m-1}+q^{\frac{m+2}2}-q^{\frac{m}2}}2$&$\frac{(q^m-1)(q^{\frac{m-2}2}-1)}{q^2-1}$\\
            $\frac{q^m-1}2$&$q-1$\\
            \bottomrule
        \end{tabular}
    \end{center}
    \label{Tab12}
\end{table}

\begin{theorem}\label{th8}
    The BCH code $\mathcal{C}_{(q,m,2,\delta_2)}$ has parameters $[\frac{q^m-1}2,k,\delta_2]$, where
    \begin{itemize}
        \item [{\rm (\romannumeral1)}] if $m$ is odd, then $k=2m+1$ and the weight distribution of $\mathcal{C}_{(q,m,2,\delta_2)}$ is listed in Table \ref{Tab11}.
        \item [{\rm (\romannumeral2)}] if $m$ is even, then $k=\frac{3m}2+1$ and  the weight distribution of $\mathcal{C}_{(q,m,2,\delta_2)}$ is listed in Table \ref{Tab12}.
    \end{itemize}
\end{theorem}

\begin{IEEEproof}
    Let $\alpha$ be a primitive element of $\mathbb{F}_{q^m}$, then $\alpha^2$ is a primitive $n$-th root of unity in $\mathbb{F}_{q^m}$. From Lemmas \ref{l9} and \ref{l10}, the code $\mathcal{C}_{(q,m,2,\delta_2)}$ has three nonzeros, and $1$, $\alpha^{2\delta_1}$ and $\alpha^{2\delta_2}$ are three non-conjugate roots of its parity-check polynomial. The dimension of $\mathcal{C}_{(q,m,2,\delta_2)}$ follows from Lemmas \ref{l9} and \ref{l10}.
   
    Case 1. $m$ is odd. Similar to the above discussion, $\mathcal{C}_{(q,m,2,\delta_2)}$ has the same weight distribution with the code
    \[\begin{split}
    \left\lbrace v_4(a,b,c): a,b\in \mathbb{F}_{q^m}, c\in \mathbb{F}_q\right\rbrace,
    \end{split}\]
    where 
    \[\begin{split}
    v_4(a,b,c)=\left( \text{Tr}_q^{q^m}( a\alpha^{(q^{\frac{m-1}2}+1)\ell}+b\alpha^{(q^{\frac{m-3}2}+1)\ell })+c \right)_{\ell=0}^{n-1}.
    \end{split}\]
    
    When $c=0$, the weight distribution of $v_4(a,b,c)$ is determined in Theorem \ref{th6}. When $c\neq 0$, we have
    \[\begin{split}
     &w(v_4(a,b,c))\\
     =&n-\sum_{\ell=0}^{n-1}\frac{1}q\sum_{y\in \mathbb{F}_q}\zeta_p^{\text{Tr}^{q}_p( y\text{Tr}_q^{q^m}( a\alpha^{(q^{\frac{m-1}2}+1 )\ell}+b\alpha^{(q^{\frac{m-3}2}+1)\ell })+yc) }\\
     =&n-\frac{1}q\sum_{y\in \mathbb{F}_q} \zeta_p^{\text{Tr}^{q}_p(yc) }\sum_{\ell=0}^{n-1}\zeta_p^{\text{Tr}^{q}_p( y\text{Tr}_q^{q^m}( a\alpha^{(q^{\frac{m-1}2}+1 )\ell }+b\alpha^{(q^{\frac{m-3}2}+1)\ell })) }\\
     =&n-\frac{1}{2q}\sum_{y\in \mathbb{F}_q} \zeta_p^{\text{Tr}^{q}_p(yc) }\sum_{\ell=0}^{2n-1}\zeta_p^{\text{Tr}^{q}_p( y\text{Tr}_q^{q^m}( a\alpha^{(q^{\frac{m-1}2}+1 )\ell }+b\alpha^{(q^{\frac{m-3}2}+1)\ell })) }\\
     =&n-\frac{1}{2q}\sum_{y\in \mathbb{F}_q} \zeta_p^{\text{Tr}^{q}_p(yc) }\sum_{x\in \Fqm^*}\zeta_p^{\text{Tr}^{q}_p(yQ_{a,b}(x)) }\\
     =&n-\frac{1}{2q}\sum_{y\in \mathbb{F}_q} \zeta_p^{\text{Tr}^{q}_p(yc) }\sum_{x\in \Fqm}\zeta_p^{\text{Tr}^{q}_p(yQ_{a,b}(x)) }\\
      =&\frac{q^m-q^{m-1}-1}2-\frac{1}{2q}\sum_{y\in \mathbb{F}_q^*} \zeta_p^{\text{Tr}^{q}_p(yc) }\sum_{x\in \Fqm}\zeta_p^{\text{Tr}^{q}_p(yQ_{a,b}(x)) },
    \end{split}\]
    where $Q_{a,b}(x)={\rm Tr}_q^{q^m}( ax^{q^{\frac{m-1}2}+1 }+bx^{q^{\frac{m-3}2}+1})$. Let $\eta$ be the quadratic character of $\mathbb{F}_q$, $r_{a,b}$ be the rank of $Q_{a,b}(x)$, and $T(a,b)=\sum_{x\in \Fqm}\zeta_p^{\text{Tr}^{q}_p(Q_{a,b}(x)) }$. From Lemma \ref{l4}, 
    \[\begin{split}
    \sum_{x\in \Fqm}\zeta_p^{\text{Tr}^{q}_p(yQ_{a,b}(x)) }=\eta(y^{r_{a,b}})T(a,b).
    \end{split}\]
     It follows that the weight of codeword $v_4(a,b,c)$ is  
    \[\begin{split}
    \frac{q^m-q^{m-1}-1}2-\frac{T(a,b)}{2q}\sum_{y\in \mathbb{F}_q^*} \zeta_p^{\text{Tr}^{q}_p(yc) }\eta(y^{r_{a,b}}).
     \end{split}\]
    
   There are two cases.
    
    If $r_{a,b}$ is even, then
    $$  w(v_4(a,b,c))=\frac{q^m-q^{m-1}-1}{2}+\frac{T(a,b)}{2q}.$$
    
    If $r_{a,b}$ is odd, then
    \begin{align*}
    w(v_4(a,b,c))&=\frac{q^m-q^{m-1}-1}{2}-\frac{T(a,b)}{2q}\sum_{y\in \mathbb{F}_q^*}\zeta_p^{{\rm Tr}_p^q(yc)}\eta(y)\\
    &=\frac{q^m-q^{m-1}-1}{2}-\frac{\eta(c)T(a,b)G_q}{2q}.
    \end{align*}
    Combining Lemmas \ref{l4} and \ref{l14}, the desired conclusion on the weight distribution then follows.

    Case 2. $m$ is even. Similar to Case 1, $\mathcal{C}_{(q,m,2,\delta_2)}$ has the same weight distribution with the code
    \[\begin{split}
    \left\lbrace v_5(a,b,c): a\in \mathbb{F}_{q^{\frac{m}2}}, b \in \mathbb{F}_{q^m},c\in \mathbb{F}_q\right\rbrace,
    \end{split}\]
    where 
    \[\begin{split}
    v_5(a,b,c)=\left( {\rm Tr}_q^{q^{\frac{m}2}}(a\alpha^{(q^{\frac{m}2}+1)\ell}) +{\rm Tr}_q^{q^{m}}( b\alpha^{(q^{\frac{m-2}2}+1)\ell }) +c\right)_{\ell=0}^{n-1}.
    \end{split}\]
    
    When $c=0$, the weight distribution of $v_5(a,b,c)$ is determined in Theorem \ref{th6}. When $c\neq 0$, similar to Case 1, the weight of codeword $v_5(a,b,c)$ is
    \[\begin{split}
    \frac{q^m-q^{m-1}-1}{2}-\frac{\overline{T}(a,b)}{2q}\sum_{y\in \mathbb{F}_q^*}\zeta_p^{{\rm Tr}_p^q(yc)}\eta(y^{r_{a,b}}),
    \end{split} \]
    where $\overline{T}(a,b)=\sum_{x\in \Fqm} \zeta_p^{{\rm Tr}_p^q(\overline{Q}_{a,b}(x))}$ and 
    \[\begin{split}
    \overline{Q}_{a,b}(x)= {\rm Tr}_q^{q^{\frac{m}2}}( ax^{q^{\frac{m}2}+1}) +{\rm Tr}_q^{q^m}( bx^{q^{\frac{m-2}2}+1}).
    \end{split}\] 
   Thanks to \cite{luo2009exponential}, the value distribution of $\overline{T}(a,b)$ is already known which is presented in Table \ref{Tab5}. There are two cases.
    
    If $r_{a,b}$ is even, then
    $$  w(v_5(a,b,c))=\frac{q^m-q^{m-1}-1}{2}+\frac{\overline{T}(a,b)}{2q}.$$
    
    If $r_{a,b}$ is odd, then
    \begin{align*}
    w(v_5(a,b,c))=\frac{q^m-q^{m-1}-1}{2}-\frac{\eta(c)\overline{T}(a,b)G_q}{2q}.
    \end{align*}
    
    Thus, the desired conclusion on the weight distribution then follows.
\end{IEEEproof}\

\begin{example}
    When $(q,m)=(3,3)$, the BCH code $\mathcal{C}_{(q,m,2,\delta_2)}$ is a $[13,7,4]$ code over $\mathbb{F}_3$ with weight enumerator $1+26z^{4}+156z^{6}+624z^{7}+494z^{9}+780z^{10}+78z^{12}+28z^{13}$. The best known linear code over $\mathbb{F}_3$ with length $13$ and dimension $7$ has minimum distance $5$ in the Datebase.
\end{example}

\begin{example}
    When $(q,m)=(3,4)$, the BCH code $\mathcal{C}_{(q,m,2,\delta_2)}$ is a $[40,7,22]$ code over $\mathbb{F}_3$ with weight enumerator $1+280z^{22}+300z^{24}+336z^{25}+240z^{27}+600z^{28}+168z^{30}+240z^{31}+20z^{36}+2z^{40}$. This code has the same parameters with the best known in the Datebase.
\end{example}

\subsection{The weight distribution of BCH codes of length $(q^m-1)/(q-1)$}

   In this subsection, we study the weight distribution of BCH codes of length $n=(q^m-1)/(q-1)$, where $m\geq q$. 
   
  \begin{lemma}\label{l15}
  	Let $q>3$ and $i$ be an integer with $1\leq i\leq n-1$. Denote the $q$-adic expansion of $i$ by $\sum_{\ell=0}^{m-1}i_{\ell}q^{\ell}$. If $i$ is a $q$-cyclotomic coset leader modulo $n$, then $i_{m-1}=0$. Suppose $m-1=a(q-1)+b$, where $a\geq 1$ and $0\leq b\leq q-2$ are integers. Let $\epsilon=a+1$ when $b=q-2$ and $\epsilon=a$ when $0\leq b\leq q-3$. If $i_{\ell}=q-1$ for all $m-1-\epsilon \leq \ell \leq m-2$, then $1\leq i_{\ell-1}\leq i_{\ell}$ for all $1\leq \ell \leq  m-2$.
  \end{lemma}
  
  \begin{IEEEproof}
  	The first statement of this lemma comes from Lemma \ref{l8}. For every positive integer $\mu$, if $i_{\mu}\neq 0$, we have $i_{\mu-1}\leq i_{\mu}$. Otherwise, we have $[iq^{m-1-\mu}]_n<i$, a contradiction. It follows that $(i_0,i_1,\ldots,i_{m-1})$ must be of the form $(I_0,I_1,\ldots,I_{v})$, where $v$ is some non-negative integer and
  	\[\begin{split}
  	I_e=(\overbrace{1\ldots 1}^{n_{e,1} }\overbrace{2\ldots 2}^{n_{e,2}}\ldots \overbrace{q-1\ldots q-1}^{n_{e,q-1}} 0),~n_{e,f}\geq 0,
  	\end{split}\] 
  	for every $0\leq e\leq v$. We now prove $v=0$. Let $\kappa=\sum_{f=1}^{q-1}n_{0,f}$, from $[iq^{m-1-\kappa}]_n\geq i$, we have $n_{0,q-1}\geq n_{v,q-1}$. Similarly, we have $n_{e,q-1}\geq n_{v,q-1}$ for all $0\leq e \leq v$. Note that
  	\[\begin{split}
  	[iq]_n=2+\sum_{\ell=1}^{m-n_{v,q-1}-1}(i_{\ell-1}+1)q^\ell+\sum_{m-n_{v,q-1}+1}^{m-1}q^{\ell}.
  	\end{split} \]
  	Denote the $q$-adic expansion of $[iq^{n_{v,q-1}}]_n$ by $\sum_{\ell=0}^{m-1}i'_{\ell}q^{\ell}$, then $(i'_0,i'_1,\ldots, i'_{m-1 })$ must be of the form $(I'_0,I'_1,\ldots,I'_{v})$, where  
  	\[\begin{split}
  	I_0'=(\overbrace{1\ldots 1}^{n_{v,(q-1)}-1 }\overbrace{2\ldots 2}^{n_{0,1}+1}\overbrace{3\ldots 3}^{n_{0,2}}\ldots \overbrace{q-1\ldots q-1}^{n_{0,q-2}}0),
  	\end{split}\]
  	\[\begin{split}
  	I_e'=(\overbrace{1\ldots 1}^{n_{e-1,q-1}-1 }\overbrace{2\ldots 2}^{n_{e,1}+1}\overbrace{3\ldots 3}^{n_{e,2}}\ldots \overbrace{q-1\ldots q-1}^{n_{e,q-2}}0),
  	\end{split}\]
  	for every $1\leq e\leq v$. It follows that $n_{e,q-2}\geq n_{v,q-1}$ for all $0\leq e \leq v$. By the same way, we have $n_{e,f}\geq n_{v,q-1}$ for all $0\leq e \leq v$ and $2\leq f \leq q-2 $, and $n_{e,1}\geq n_{v,q-1}-1$ for all $0\leq e\leq v$. Therefore,
  	\[\begin{split}
  	m&=\sum_{e=0}^{v} \left( \sum_{f=1}^{q-1}n_{e,f}+1\right)\\
  	&\geq (v+1)(q-1)n_{v,q-1}\\
  	&\geq (v+1)(q-1)\epsilon,
  	\end{split}\]
  	since $n_{v,q-1}\geq \epsilon$. If $v\geq 1$, we have $a(q-1)+b+1\geq 2(q-1)\epsilon$, a contradiction.
  \end{IEEEproof}
  
  \begin{lemma}\label{l16}
  	Let $q>3$ be a prime power and $m\geq q$. Suppose $m-1=a(q-1)+b$, where $a\geq 1$, $0\leq b\leq q-2$ are integers.
  	\begin{itemize}
  		\item [(\romannumeral1)]If $b=0$, i.e., $m=a(q-1)+1$, then the first largest $q$-cyclotomic coset leader modulo $\frac{q^m-1}{q-1}$ is 
  		\[\delta= \frac{q^m-1-q^{m-1}-\sum_{\ell=1}^{q-2}q^{a\ell}}{q-1} \]
  		and $|\mathbb{C}_{\delta}|=m$.
  		\item [(\romannumeral2)]If $b=1$, i.e., $m=a(q-1)+2$, let $A=\lfloor \frac{q-1}2 \rfloor$, then the first largest $q$-cyclotomic coset leader modulo $\frac{q^m-1}{q-1}$ is 
  		\[\delta=\frac{q^{m}-1-q^{m-1}-\sum_{\ell=1}^{A}q^{a\ell}-\sum_{\ell=A+1}^{q-2}q^{a\ell+1}}{q-1}. \]
  		Moreover, $|\mathbb{C}_{\delta}|=\frac{m}2$ when $q$ is odd, and $|\mathbb{C}_{\delta}|=m$ when $q$ ie even.
  		\item [(\romannumeral3)]If $b=q-2$, i.e., $m=(a+1)(q-1)$, then the first largest $q$-cyclotomic coset leader modulo $\frac{q^m-1}{q-1}$ is 
  		\[\delta= \frac{q^{m}-1-q^{m-1}-\sum_{\ell=1}^{q-2}q^{(a+1)\ell-1}}{q-1} \]
  		and $|\mathbb{C}_{\delta}|=a+1$.
  	\end{itemize}
  \end{lemma}
  
  \begin{IEEEproof}
  	We just give the proof for Case (\romannumeral2), since the proofs in the other cases are similar. Clearly, the $q$-adic expansion of $\delta$ is of the form  
  	\[\begin{split}
  	\sum_{i=1}^{A} \sum_{\ell=(i-1)a}^{ia-1} iq^{\ell}+\sum_{\ell=Aa}^{(A+1)a}(A+1)q^{\ell} +\sum_{i=A+2}^{q-1} \sum_{\ell=(i-1)a+1}^{ia} iq^{\ell},
  	\end{split} \]
  	and it is easy to check that $\delta$ is a $q$-cyclotomic coset leader modulo $n$. Moreover, $|\mathbb{C}_{\delta}|=\frac{m}2$ if $q$ is odd, and $|\mathbb{C}_{\delta}|=m$ if $q$ is even. 
  	
  	We now prove that $\delta$ is the largest integer in the set of all coset leaders. Suppose there is an integer $s$ with $\delta<s<n$ which is a $q$-cyclotomic coset leader modulo $n$ and the $q$-adic expansion of $s$ is $\sum_{\ell=0}^{m-1}s_{\ell}q^{\ell}$. By Lemma \ref{l15}, $(s_0,s_1,\ldots, s_{m-1 })$ must be of the form
  	\[\begin{split}
  	(\overbrace{1\ldots 1}^{n_{1} }\overbrace{2\ldots 2}^{n_{2}}\ldots \overbrace{q-1\ldots q-1}^{n_{q-1}} 0),
  	\end{split}\]
  	where $n_{q-1}\geq a$, $n_{\ell}\geq n_{q-1}$ for $2\leq \ell\leq q-2$, and $n_{1}\geq n_{q-1}-1$. Firstly, $n_{q-1}=a$. Otherwise, 
  	\[\begin{split}
  	m-1&=\sum_{\ell=1}^{q-1}n_{\ell}\geq(q-1)n_{q-1}-1\\
  	&\geq a(q-1)+q-2> m-1,
  	\end{split}\]
  	a contradiction. Secondly, $n_{\ell}\leq a+1$ for all $2\leq \ell \leq q-2$. Otherwise, there is an integer $v$ such that $n_{v}=a+2$, then we have
  	\[\begin{split}
  	[sq^{(q-v)a+2}]_n=\sum_{\ell=0}^{a}q^{\ell}+\sum_{i=2}^{q-1} \sum_{\ell=(i-1)a+1}^{ia} iq^{\ell}<\delta,
  	\end{split} \]
  	a contradiction. Thirdly, $n_1\geq a$. Otherwise, suppose $n_1=a-1$, then there are two integers $2\leq u<v\leq q-2$ such that $n_{u}=n_{v}=a+1$. It is easy to check that 
  	\[\begin{split}
  	[sq^{(q-u)a+1}]_n<s,
  	\end{split}\] 
  	a contradiction. Therefore, $n_0=a$ and there is an integer $v$ with $A+2\leq v\leq q-2$ such that $n_v=a+1$. From $[sq^{(q-v)a+1}]_n\geq s$, we have $2v\leq q+1$, a contradiction.
  	
  	Collecting all the conclusions above, we conclude that $\delta$ is the largest coset leader.
  \end{IEEEproof}

  Based on Lemma \ref{l16}, we calculate the weight distribution of BCH code $\widehat{\cC}_{(q,m,q-1,\delta)}$ as follows.
  
 \begin{theorem}
 	Let $q>3$ be a prime power and $m\geq q$. Suppose $m-1=a(q-1)+b$, where $a\geq 1$ and $0\leq b\leq q-2$ are integers. 
 	\begin{itemize}
 		\item [(\romannumeral1)]If $b=0$ or $b=1$ and $q$ is even, then the BCH code $\widehat{\cC}_{(q,m,q-1,\delta)}$ is a $[\frac{q^m-1}{q-1}, m, q^{m-1}]$ one-weight code. 
 		\item [(\romannumeral2)]If $b=1$ and $q$ is odd, then the BCH code $\widehat{\cC}_{(q,m,q-1,\delta)}$ is a $[\frac{q^m-1}{q-1},\frac{m}2, (q^{\frac{m}2}+1)q^{\frac{m}2-1}]$ one-weight code. 
 	\end{itemize}
 \end{theorem}
 
 \begin{IEEEproof}
 		We just give the proof for Case (\romannumeral2), since the proofs in the other cases are similar. Let $\alpha$ be a primitive element of $\mathbb{F}_{q^m}$, then $\alpha^{q-1}$ is a primitive $n$-th root of unity in $\mathbb{F}_{q^m}$. From Lemma \ref{l16}, the BCH code $\widehat{\cC}_{(q,m,q-1,\delta)}$ has one nonzero and $\alpha^{(q-1)\delta}$ is a root of its parity-check polynomial. Let $\tau=q^{m-1}+\sum_{\ell=1}^{\frac{q-1}2}q^{a\ell}+\sum_{\ell=\frac{q+1}2}^{q-2}q^{a\ell+1}$, then $-(q-1)\delta\equiv \tau ~({\rm mod}~q^m-1)$. By Lemma \ref{l3}, 
 		\[\begin{split}
 		\widehat{\cC}_{(q,m,q-1,\delta)}=\left\lbrace \left(\text{Tr}^{q^{\frac{m}2}}_q(a\alpha^{\tau\ell})\right)_{\ell=0}^{n-1} : a\in \ff_{q^{\frac{m}2}}\right\rbrace.
 		\end{split}\]
 		 Let $\beta=\alpha^{q^{\frac{m}2}+1}$. Since $\text{Tr}^{q^{\frac{m}2}}_q(a\alpha^{\tau\ell})=\text{Tr}^{q^{\frac{m}2}}_q(a^q\alpha^{q\tau\ell})$ and $\gcd(q\tau, q^m-1)=\frac{(q-1)(q^{\frac{m}2}+1)}2$, it follows that the BCH code $\widehat{\cC}_{(q,m,q-1,\delta)}$ has the same weight distribution with the following code
 		\[\begin{split}
 		\left\lbrace c(a)=\left(\text{Tr}^{q^{\frac{m}2}}_q(a\beta^{\frac{q-1}2\ell})\right)_{\ell=0}^{n-1}: a\in \ff_{q^{\frac{m}2}} \right\rbrace.
 		\end{split}\]
 		Let $n'=\frac{2(q^{\frac{m}2}-1)}{q-1}$ and 
 		\[\begin{split}
 		C'=\left\lbrace c'(a)=\left(\text{Tr}^{q^{\frac{m}2}}_q(a\beta^{\frac{q-1}2\ell})\right)_{\ell=0}^{n'-1}: a\in \ff_{q^{\frac{m}2}} \right\rbrace.
 		\end{split}\]
 		Note that $\gcd(\frac{q^{\frac{m}2}-1}{q-1},\frac{q-1}2)=1$. From \cite[Theorem 15]{ding2013irreducible}, $C'$ is a $[n',\frac{m}2, 2q^{\frac{m}2-1}]$ one-weight code over $\mathbb{F}_q$. It is easy to check that
       \[\begin{split}
       c(a)=\overbrace{c'(a)\parallel \cdots \parallel c'(a)}^{\frac{n}{n'}}.
       \end{split} \]
	  Hence, $C$ is a $[n,\frac{m}2, (q^{\frac{m}2}+1)q^{\frac{m}2-1}]$ one-weight code over $\mathbb{F}_q$.
 \end{IEEEproof}

\section{Conclusion} \label{sec5}

The dimension of narrow-sense BCH codes of length $\frac{q^m-1}\lambda$ over $\mathbb{F}_q$ has been obtained, where $\lambda$ is a positive divisor of $q-1$. For the case $\lambda=1$ and $q-1$, the dimension of $\cC_{(q,m,\lambda,\delta)}$ was determined in \cite{li2017lcd,liu2017dimensions}. For the case $\lambda=q-1$ and $m$ is even, the dimension of $\cC_{(q,m,\lambda,\delta)}$ with designed distance $\delta$ with $2\leq \delta \leq q^{\frac{m}2}$ was settled in \cite{li2017lcd}. We settled its dimension for all $\delta$ with $2\leq \delta\leq \frac{q^{\frac{m+2}2}-1}{q-1}$. For $\lambda=2$ and $q-1$, the weight distribution of $\cC_{(q,m,\lambda,\delta)}$ was studied. We find the first few largest $q$-coset leaders modulo $n=\frac{q^m-1}2$ and a trace representation for the codewords in $\mathcal{C}_{(q,m,2,\delta_i)}$ and $\widehat{\mathcal{C}}_{(q,m,2,\delta_i)}$ for $i=1,2$. In addition, by using exponential sums and the theory of quadratic forms over finite fields, the weight distribution of $\mathcal{C}_{(q,m,2,\delta_i)}$ and $\widehat{\mathcal{C}}_{(q,m,2,\delta_i)}$ was determined. Moreover, the first largest $q$-coset leader modulo $\frac{q^m-1}{q-1}$ was determined for three special cases, and the weight distribution of a class of BCH codes of length $\frac{q^m-1}{q-1}$ was also determined. A class of BCH codes meeting the Griesmer bound has been given. These results generalized those from \cite{li2017lcd,liu2017dimensions,li2016narrow}.

\section*{Acknowledgements}

The authors wish to express their gratitude to Prof. Vladimir Sidorenko, the Associate Editor, and three anonymous reviewers who gave many helpful comments and suggestions to greatly improve the presentation of the paper.

\bibliographystyle{IEEEtran}

\end{document}